\documentclass[aps,pre,twocolumn,groupedaddress]{revtex4-1}
\usepackage{lipsum}
\usepackage{graphicx}
\usepackage{amsmath,amssymb}
\usepackage{bm}
\usepackage{dcolumn}
\usepackage{theorem}
\usepackage{wrapfig}
 \usepackage{mathtools}
\usepackage{xcolor}
 \usepackage{txfonts}
\mathtoolsset{centercolon}
\usepackage{amsmath}
\usepackage{graphicx}
\usepackage{bm}
\usepackage{float}
\usepackage{epstopdf}
\makeatletter

\begin{document}

\title{
\textit{\footnotesize	Physical Review E\  97, 062152 (2018)}\\ \textit{\footnotesize DOI: 10.1103/PhysRevE.97.062152} \\   \vskip0.5cm Fluid Mixtures in Nanotubes.\\
% Application to ethanol-water mixtures\\
%in carbon nanotubes}
}

\author{Henri Gouin$^{1\dagger}$, Augusto Muracchini$^2$, and Tommaso Ruggeri$^2$}

%\thanks{}
\affiliation{
	$^1$Aix Marseille Univ, CNRS,  
	IUSTI \, UMR 7343, 13453 Marseille, France\\
$^{2}$Department of Mathematics, University of Bologna, 40123 Bologna, Italy  }

\date{April 24, 2018}
\email[]{$\dagger$ Author for correspondence: Henri Gouin\\ henri.gouin@univ-amu.fr; henri.gouin@yahoo.fr,\\ Other E-mails: \\ augusto.muracchini@unibo.it, tommaso.ruggeri@unibo.it}

\begin{abstract}
  The aim of the paper is the study  of
 fluid mixtures in nanotubes by the   methods of continuum mechanics.  The model starts from a statistical distribution in mean-field molecular theory and uses a density expansion of
 Taylor series. We get a  continuous expression of the   volume   free energy  with density's spatial-derivatives limited at the second order.
The nanotubes can be filled with liquid or vapor according to the chemical characteristics of   the walls and of liquid 
 or vapor mixture-bulks.  An example of two-fluid mixture constituted of water and ethanol inside carbon nanotubes at $20^\circ $ C  is considered.
When    diameters are small enough,  nanotubes are   filled with liquid-mixture whatever are
 the liquid or vapor mixture-bulks. The carbon wall
 influences the ratio of the fluid components in favor of ethanol.   The fluid-mixture
 flows across nanotubes can be much more important than classical
 ones and if the external  bulk  is vapor, the flow can be  
 several hundred thousand times larger than Poiseuille flow.\\

\noindent PACS Numbers: 61.46.Fg, 61.20.Gy, 68.35.Md, 47.61.-k \\
\noindent  Keywords: {nanotubes;  fluid mixtures; nanotube flows,
	Fluid-mixture/solid interactions.} 
\end{abstract}
  \maketitle

%\titlerunning{Fluid binary mixtures in natubes}        % if too long for running head


%  $^{(b)}$ Department of Mathematics and Research Center of Applied Mathematics, University
%of Bologna, Via Saragozza 8, 40123 Bologna, Italy}

\section{Introduction}

The technical development of sciences allows us to observe phenomena at
length scales of a very few number of nanometers. The observations  reveal new
behaviors that are often surprising and essentially different from those
usually observed at a microscopic or macroscopic scales \cite{Harris,Gouin 6}. 
Experiments prove that   liquid densities   change in very narrow
pores \cite{Ball} and the conventional laws of capillarity are disqualified
when they are applied to fluids confined inside porous materials \cite{Bear}.
Iijima, the discoverer of carbon nanotubes \cite{Iijima}, was fascinated
by Kr\"atschmer \emph{et al}'s paper \cite{Kr} and decided to launch out
into a detailed study of nano-materials. Since the late 1900s the literature
has become abundant regarding  technology and flows inside nanotubes \cite%
{Mattia}. Nevertheless, simple models proposing qualitative behaviors need
to be developed. \newline 
In this paper, our aim is to investigate an example of
mixture-solid interaction in statics as well as in dynamical conditions by
using the   methods of continuum mechanics. These continuous methods are experimentally realistic until nanotube diameter sizes of a very small number of   nanometers \cite{Bocquet} \newline
To propose an analytic expression of   densities for mixture-films of
nanometric thicknesses near a smooth solid wall, we add an interaction
energy at the solid wall to a density-functional which represents the volume  free 
energy of the   mixture \cite{Gouin 3, Gouin 4,Gouin5}. These two
energies are obtained by series expansions using London's potentials for
fluid-fluid and fluid-solid interactions \cite{Gouin 2}. The  
functional is extended from studies by van der Waals, Cahn and Hilliard and
many others \cite{van der Waals,Cahn,Rowlinson,Widom}.\newline

\noindent We consider a nanotube made up of a cylindrical hollow tube whose
diameter is of a few number of nanometers. The length of the nanotube is of
microscopic size and the cylinder wall is solid   \cite{Harris}. The nanotube is immersed in an homogeneous liquid or vapor bulk made up with two fluids which fill the interior of the nanotube.

\noindent In nanofluidics, the interactions between fluids and
solid walls can dominate over the hydrodynamic behaviors. The
mixture compressibility near a solid wall is extremely important
\cite{Ball}. In liquid or vapor bulks, we express the
chemical potentials of fluid-components with only a first-order
development taking account of   isothermal sound-velocities of
single bulks \cite{Gouin 6}. The equations of equilibrium and
motion of fluid mixtures inside the nanotube take  the
fluid super-deformations into account \cite{Gouin 3}. In cylindrical
representation, two differential equations  are obtained and   the profiles of the fluid-mixture
densities in the cylinder can be deduced. \newline

\noindent The results are applied to carbon nanotubes filled with a mixture
constituted of water and ethanol and extend those obtained for simple fluids in \cite%
{Gouin9,GS}. The nanotube diameter ranges from 1
 to 100  nanometers. Due to energetic properties, the case
of liquid and vapor separated by an interface inside the nanotube is not possible when the
nanotube diameter is small enough. The mixture inside nanotubes is liquid
and the ratio between water and ethanol   significantly changes from the
mixture-bulk ratio. Recently, it was shown, by using non-equilibrium
molecular dynamics simulations, that liquid flows through a membrane
composed of an array of aligned carbon nanotubes are a lot faster than it
would be predicted by conventional fluid-flow theory \cite{Majumder}.
These high   velocities are possible because of a frictionless surface
at the nanotube wall \cite{Ma}. By calculating the variation of  
viscosity and slip length as a function of the nanotube  diameters, the
results can be fully explained in the context of continuum fluid mechanics
\cite{Thomas}. In our model, we find that a spectacular effect must appear
for tiny carbon nanotubes when the mixture bulk outside the nanotube is
constituted of vapor: the mixture flows through the nanotube can be multiplied
by a factor of several hundred thousand times what is found for Poiseuille's
model. \newline

The paper is organized as follows:\newline In   Section 2, thanks
to the mean-field theory with hard-sphere molecules, we present
a continuous form of   free  energy for inhomogeneous mixtures.  In Section 3,
the equations of equilibrium and boundary conditions at a solid
wall are written. In Section 4, we consider the chemical potentials of
fluid components near a phase and in Section 5, we study the special
case of ethanol and water mixture in carbon nanotubes. Thanks to
Hamaker's constants, we can determine the
 profiles of densities at equilibrium. When the pressure of vapor
 bulk is not too important with respect to the liquid bulk pressure, the
 carbon nanotube is filled by liquid mixture. Numerical
 computations yield the densities profiles of water and ethanol  
 inside the nanotubes. In Section 6, the results are extended in the motion case and, as it is experimentally verified,  with a classical viscosity. A conclusion ends the paper.

\section{\label{sec2} A second-gradient fluid-mixture energy}
 {In a single fluid where the density is not uniform, the intermolecular forces exert on a given molecule a resultant force accounting for the effects of surface tension \cite{Rocard}. The force derives from a system of pressures and stresses that differs from the classical isotropic pressure, which assumes uniformity of density. The kinetic theory of the gases makes it possible to obtain this system of tension and allows us to study   the equilibrium of the fluid by layers of equal density and produces a good modelization of the superficial tension.
It has been known for a long time, that the law of statistical distribution of molecules around one of them is almost inaccessible even for a distribution of uniform density \cite{Ornstein}. If we take a regularized law of distribution at a distance of a given molecule, then we ignore density fluctuations, but the effect of fluctuations tends to be eliminated by integration. This is why, it seems to us possible to approach the problem of the field of forces without presupposing the problem solved and without knowing the true distribution of the molecules. An important property is that the probability of presence of a molecule at a distance less than the molecular diameter is zero. 
The potential of intermolecular forces is a rapidly decreasing function of distance. It is enough to know the law of distribution at the effective distance and thus the fluctuations of the law  beyond this distance no longer influence the result. The case corresponds to the reality given that the preponderant forces are of the type  $1/ r ^ 7$. It is then possible to develop in series the distribution gap of the molecules. An energy density is then expressed as a measure per unit volume and allows a description in mechanics of  continuous media.
Such an energy model allows us to obtain a good variation of density in liquid-vapor interfaces even if they are spherical and of  nanometer size \cite{Isola2}. It is even possible to get variation of fluid density in the vicinity of a solid wall. As it is proposed in \cite{Gouin 2}, we can extend the model for inhomogeneous fluid mixtures. 
Then, it is possible to propose analytic expressions for the interaction potentials in terms of densities of the two species of a mixture and consequently to get a continuum mechanics description of volume energy which is valid at nanometer size of interfaces.}

\subsection{\label{subsecA}Energy per unit volume}
In the mean-field theory with hard-sphere molecules and for each  
constituent  of a fluid-mixture, the fluid's molecules
are identical. The central forces between molecules of
constituent $i$,\ $i \in \left\lbrace1,2 \right\rbrace$ derive from a potential  {$\,\varphi _{i}(r)$ and between
molecules of the two constituents from a potential $\,\varphi _{3}(r)$,
where $r$ is the distance between the centers of the two molecules} \cite%
{Lifshitz}.\newline
In three-dimensional Euclidean medium $\mathcal{D}$, the potential energy $%
W_{{o1}}$ resulting from the combined action of all the molecules on the
molecule of constituent ${1}$, located at origin $O$, is assumed to be
additive such that {
\begin{equation*}
W_{{o1}}=\sum_{N_1}\,\varphi _{1}(r)+\sum_{N_2}\,\varphi
_{3}(r),
\end{equation*}
}\noindent where $N_{1}$  denotes all molecules of constituent $1$ (except
for the molecule located at origin $O$) and   $N_{2}$ of constituent 2, respectively;  $r$ denotes the distance of  molecules to the molecule of constituent $1$
located at $O$.
With similar notations, we get {
\begin{equation*}
W_{{o2}}=\sum_{N_2}\,\varphi _{2}(r)+\sum_{N_1}\,\varphi
_{3}(r).
\end{equation*}}The number of molecules of constituent $i$ in   volume $dv
$ is represented by { $n_{i}(x,y,z)\,dv$}, where $dv$ denotes the volume
element in $\mathcal{D}$ at point of coordinates $x,y,z$, and in a
continuous representation,
{
\begin{equation*}
W_{{o1}}=\iiint_{\mathcal{D}}\,\varphi _{1}(r)\,n_{1}\,dv+\iiint_{%
\mathcal{D}}\,\varphi _{3}(r)\, n_{2}\,dv,
\end{equation*}%
}
with
{
\begin{equation*}
\iiint_{\mathcal{D}}\,\varphi _{1}(r)\,n _{1}\,dv=\int_{\sigma
_{1}}^{\infty }\,\varphi _{1}(r)\left[ \iint_{S(r)}n _{1}\,ds\right]
\,dr,
\end{equation*}
}%
and
{
\begin{eqnarray*}
\iiint_{\mathcal{D}}\,\varphi _{3}(r)\,n _{2}\,dv&=&\\ \int_{\frac{1}{2}%
\left( \sigma _{l}+\sigma _{2}\right) }^{\ \ \infty }\,\varphi _{3}(r)%
&&\left[ \iint_{S(r)}n _{2}\,ds\right] \,dr,
\end{eqnarray*} 
}where \textit{ds} is the measure of area, $S(r)$ is the sphere of center $O$ and
radius $r$, and $\sigma _{i}$ is the molecular diameter of molecules of
component $i,\, i\in \{1,2\}$. {We assume that $n _{i}(x,y,z),\ i\in \{1,2\}$
are analytic functions of coordinates $x,y,z$, {\it i.e.}  
\begin{align*}
\begin{split}
&n _{i}=n _{i}(0,0,0) \\
&+\sum_{\ell=1}^{\infty }\frac{1}{{\ell}\,!}\left[ \,x\frac{%
\partial n _{i}}{\partial x}(0,0,0)+y\frac{\partial n _{i}}{\partial y}%
(0,0,0)+z\frac{\partial n _{i}}{\partial z}(0,0,0)\,\right]^{(\ell)}.
\label{Expansion}
\end{split}
\end{align*}
}%
The method will be specially justified in two cases:\newline
$(a) $ $\varphi_k (r),\  k \in\{1,2,3\}$, are constant over a large diameter: then the distribution fluctuations of the following molecules are eliminated, this would be the case for van der Waals forces with a large radius of action.\newline
$(b) $ $\varphi_k (r), \ k \in\{1,2,3\}$, are rapidly decreasing functions of distance:  It suffices to know the law of distance distribution of molecules at the effective distance, that is, very close to one of them, and then the fluctuations of the law at a great distance, or even at a distance a little greater, no longer influence the result.\newline
In fact, the second case corresponds to reality   since the predominant forces which we will have to count with, are of type $1 / r ^ 7$.
It is then possible to develop in series the difference of $n_i$ with respect to its value in $O$ and to limit to the second order. All these comments are developed in
 \cite{Rocard,Rowlinson}. \newline We notice that for any integers $p,q,r$ we have the
relations
\begin{equation*}
\iint_{S(r)}x^{2p+1}y^{q}\,z^{r}ds=0,
\end{equation*}%
and
\begin{equation*}
\iint_{S}x^{2}ds=\iint_{S}y^{2}ds=\iint_{S}z^{2}ds=\frac{4\,\pi \,r^{4}}{3}.
\end{equation*}%
Then,
{
\begin{eqnarray*}
W_{{o1}} &=&\int_{\sigma _{1}}^{\infty }\,\varphi _{1}(r)\left[
4\,\pi \,r^{2}n {_{1o}}+\frac{2\pi }{3}\,r^{4}\Delta n{_{1o}}\right]
dr \\
&+&\int_{\frac{1}{2}\left( \sigma _{l}+\sigma _{2}\right) }^{\ \ \infty
}\,\varphi _{3}(r)\left[ 4\,\pi \,r^{2}n {_{2o}}+\frac{2\pi }{3}%
\,r^{4}\Delta n{_{2o}}\right] dr.
\end{eqnarray*}%
Here $n {_{io}}\equiv n _{i}(0,0,0)$,\ $\Delta n{_{io}}\equiv
\Delta n _{i}(0,0,0),\ i\in \{1,2\}$}  and $\Delta$ is the
Beltrami-Laplace operator. Let us denote
{
\begin{eqnarray*}
2\,m_1^2
\,k_{1} &=&\int_{\sigma _{1}}^{\infty }4\,\pi \,r^{2}\varphi
_{1}(r)\,dr,\\ 2\,m_1^2\,k_{1}\, b_{1}^{2}&=&\int_{\sigma _{1}}^{\infty }\frac{%
2\,\pi }{3}\,r^{4}\varphi _{1}(r)\,dr,  \label{lambda} \\
2\,m_1m_2\,\kappa _{3} &=&\int_{\frac{1}{2}\left( \sigma _{l}+\sigma _{2}\right)
}^{\ \ \infty }4\,\pi \,r^{2}\varphi _{3}(r)\,dr,\\ 2\,k_{3}\,m_1m_2\,
b_{3}^{2}&=&\int_{\frac{1}{2}\left( \sigma _{l}+\sigma _{2}\right) }^{\ \
\infty }\frac{2\,\pi }{3}\,r^{4}\varphi _{3}(r)\,dr,  \notag
\end{eqnarray*}
}where $\sigma_i$, $i\in \{1, 2\}$,
denotes molecular diameters of fluids,  $b_{1}$\ is the fluid's co-volume of component $1$ \cite{Rocard} and $%
b_{3}$\  is   analog to a fluid's co-volume between components $1$ and $2$.
{ To define different mass densities  $\rho_{i}$   
	of components $i$, we must introduce  the molecular masses of species   denoted $m_i$; then, $\rho_{i}=n _{{i}} m_{i} $ and
\begin{equation*}
W_{{o1}}=2\,m_{1}^{2}\,k_{1}\left[ n _{{1o}}+b_{1}^{2}\,\Delta \,n
_{{1o}}\right] +2\,m_{1}m_{2}\,k_{3}\left[ n _{{2o}}+b_{3}^{2}\,\Delta
\,n _{{2o}}\right],
\end{equation*}%
We deduce,
\begin{equation*}
W_{{o1}}=2\,m_{1}\,k_{1}\,\left[ \rho _{{1o}}+b_{1}^{2}\,\Delta \rho
_{{1o}}\right] +2\,m_{1}\,k_{3}\left[ \rho _{{2o}}+b_{3}^{2}\,\Delta
\,\rho _{{2o}}\right] ,
\end{equation*}%
where $\rho _{{io}}=n _{{io}}m_{i},\ i\in \{1,2\}$ are the mass densities
of components $i$ at $O$. Similarly, we obtain
\begin{equation*}
W_{{o2}}=2\,m_{2}\,k_{2}\,\left[ \rho _{{2o}}+b_{2}^{2}\,\Delta \rho
_{{2o}}\right] +2\,m_{2}\,k_{3}\left[ \rho _{{1o}}+b_{3}^{2}\,\Delta
\,\rho _{1o}\right] .
\end{equation*}
}We have to consider that all couples of molecules of the two components are
counted twice and the double of the potential energy density per unit volume
is
{
\begin{equation*}
2\left( n _{{1o}}\,W_{{o1}}+ n _{{2o}}\,W_{{o2}}\right) \equiv
2\left( E_{{o1}}+E_{{o2}}\right),
\end{equation*}%
with%
\begin{equation*}
\left\{
\begin{array}{c}
E_{{o1}}= n _{1o}\,m_{1}\, k_{1}\,\left[ \rho _{{1o}}+b_{1}^{2}\,\Delta
\rho _{{1o}}\right] + n _{{1o}}\,m_{1}\,k_{3}\left[ \rho
_{{2o}}+b_{3}^{2}\,\Delta \,\rho _{{2o}}\right], \\
\\
E_{{o2}}=n _{{2o}}\, m_{2}\,k_{2}\,\left[ \rho _{{2o}}+b_{2}^{2}\,\Delta
\rho _{{2o}}\right] + n _{{2o}}\,m_{2}\,\,k_{3}\left[ \rho
_{{1o}}+b_{3}^{2}\,\Delta \,\rho _{{1o}}\right] .%
\end{array}%
\right.
\end{equation*}%
}
The corresponding potential energy of the two-fluid mixture becomes
\begin{eqnarray*}
W &=&\iiint_{\mathcal{D}}\Big\{k_{1}\left[ \rho _{1}^{2}+b_{1}^{2}\,\rho
_{1}\,\Delta \rho _{1}\right] +k_{2}\left[ \rho _{2}^2+b_{2}^{2}\,\rho
_{2}\,\Delta \rho _{2}\right] \\
&&+k_{3}\left[ 2\rho _{1}\,\rho _{2}+b_{3}^{2}\,\rho _{1}\,\Delta \rho
_{2}+b_{3}^{2}\,\rho _{2}\,\Delta \rho _{1}\right] \Big\}dv.
\end{eqnarray*}%
Taking account of  
\begin{equation*}
\rho _{i}\,\Delta \rho _{j}=\rho _{i}\,\mathrm{div}\left( {\nabla }\,\rho
_{j}\right) =\mathrm{div}\left( \rho _{i}\,{\nabla }\,\rho _{j}\right) -{%
\nabla }\,\rho _{i}\,{\nabla }\,\rho _{j},
\end{equation*}%
where $\nabla$ is the gradient operator, terms $\mathrm{div}\left( \rho
_{i}\,{\nabla }\,\rho _{j}\right)$, with $i,j\in \{1,2\}$, are integrable on
the boundary of ${\mathcal{D}}$, where  $\rho _{1}$ and $\rho _{2}$ are
assumed uniform  and yield a zero contribution to $W$. We get
\begin{eqnarray}
W &=&\iiint_{\mathcal{D}}\Big\{\,k_{1}\,\rho _{1}^{2}+k_{2}\,\rho
_{2}+2\,k_{3}\,\rho _{1}\,\rho _{2}-k_{1}\,b_{1}^{2}\,\left( {\nabla }\rho
_{1}\right) ^{2}  \notag \\
&-&k_{2}\,b_{2}^{2}\,\left( {\nabla }\rho _{2}\right)
^{2}-2\,k_{3}\,b_{3}^{2}\,{\nabla }\rho _{1}\,{\nabla }\rho _{2}\Big\}\ dv.
\label{2grad}
\end{eqnarray}%
To obtain the    {volume free} energy {at given temperature $T$}, we have to take account of the kinetic
effects of molecular motions where the first terms $k_{1}\,\rho_{1}^{2}+k_{2}\,%
\rho_{2}^2+ 2\,k_{3}\,\rho _{1}\,\rho _{2}$\, in Eq. \eqref{2grad} correspond
to the total internal pressure \cite{Rocard}. Consequently, the  {volume free} 
energy reads
\begin{equation}
\varepsilon =\varepsilon _{0}(\rho _{1},\rho _{2})+\frac{1}{2}\left[%
\,\lambda _{1}\,\left( {\nabla }\rho _{1}\right) ^{2}+\lambda _{2}\,\left( {%
\nabla }\rho _{2}\right) ^{2}+2\lambda _{3}\,{\nabla }\rho _{1}\,{\nabla }%
\rho _{2}\,\right]  \label{energy capillaire},
\end{equation}%
where $\varepsilon _{0}(\rho _{1},\rho _{2})$ is the   { volume free} energy  
of the homogeneous fluid-mixture when densities are $\rho _{1}$, $\rho _{2}$ {(for the sake of simplicity, we omit to indicate $T$ in $ \varepsilon_0$)}  and
\begin{equation}
\lambda _{1}=-2\,k_{1}\,b_{1}^{2},\quad \lambda
_{2}=-2\,k_{2}\,b_{2}^{2},\quad \mathrm{and}\quad \lambda
_{3}=-2\,k_{3}\,b_{3}^{2}\,.  \label{Lambdas}
\end{equation}%
Let us note that coefficients $k_{1},k_{2},k_{3}$ are negative for
intermolecular potentials associated with attractive forces and  
coefficients $\lambda _{1},\lambda _{2},\lambda _{3}$ are positive.\\

\subsection{Energy per unit area of wall boundary}
{In 1977, John Cahn  gave simple illuminating arguments
to describe the interaction between solids and liquids \cite{Cahn2}. His model is
based on a generalized van der Waals theory of fluids treated as
attracting hard spheres \cite{Rowlinson}. It entailed assigning to the solid
surface an energy that
was a functional of the liquid density {\it at the surface}; the particular form
 of this energy is now widely known in the
literature  and was
thoroughly examined as an ad hoc approximation in a review paper by de Gennes \cite{degennes}.  
\noindent Three hypotheses are implicit in Cahn's picture,}\\
{ $(i)$ \ for the liquid density to be taken as a smooth function
  of the distance    from the solid surface,  
the correlation length is assumed to be greater
than intermolecular distances,} \\
{$(ii)$\ the forces between solid and liquid are of short range
with respect to intermolecular distances,}\\
{$(iii)$ \ the fluid  is  considered in the framework of a mean-field
theory. This
means, in particular, that the free energy of the fluid is a classical
so-called {\it gradient square functional}.}

{ The model was justified and for fluid mixtures, an extension of Cahn's model  was proposed with the same hypotheses in \cite{Gouin 2}.} The surface is a smooth solid, sharp on an atomic
scale {and  is endowed with a surface
energy per unit area of wall   $\mathcal{S}$ of domain $\mathcal{D}$. The  general form of the surface energy per unit
area  can be written
$
e=e(\rho _{1s},\rho _{2s}) 
$, 
where $\rho _{1s}$ and $\rho _{2s}$ are values of the  fluid densities at the
wall.  
We consider the case 
when the energy per unit area at the boundary is written as}
\begin{equation}
e  = -\gamma _{11}\rho _{1s}-\gamma _{21}\rho _{2s}+
\frac{1}{2}\left( \gamma _{12}\rho _{1s}^{2}+\gamma _{22}\rho
_{2s}^{2}+2\gamma _{32}\rho _{1s}\rho _{2s}\right),
\label{surfacenergy}
\end{equation}
where $\gamma _{11},\, \gamma _{21},\, \gamma _{12},\, \gamma _{22},\,$ and $\gamma _{32}$ are
positive coefficients expressing  the wall's quality with
respect to the two-fluid components of the mixture. {Values of  the coefficients have   been proposed in  \cite{Gouin 2} for the case of London's potentials and can be calculated with molecular quantities and Hamaker's constants.} \newline
 
\subsection{van der Waals' forces and Hamaker's constants}
We have introduced the intermolecular potentials associated with van der
Waals' forces in the form $\varphi (r)$.\ We consider the case of London's
forces between fluids and solid wall and we denote
{
\begin{equation*}
\varphi _{11}(r),\ \varphi _{22}(r),\ \varphi _{ss}(r),\ \varphi _{12}(r),\ \varphi _{1s}(r),\
\varphi _{2s}(r),
\end{equation*}
}the London potentials of interactions between fluid  $1$--fluid $1$,  fluid $2$--fluid $2$, solid--solid,
fluid $1$--fluid $2$, fluid $1$--solid, fluid $2$--solid,  
respectively. With the notations of Sec. \ref{subsecA} and denoting $\sigma_s$ the molecular diameter of the wall molecules, the London potentials verify
\begin{widetext}
	\begin{align*}
	\begin{split}
	&\varphi_{ii}(r) =-\frac{c_{ii}}{r^{6}},\text{ where }r>\sigma _{i}\text{ \
		and\ }\varphi _{ii}(r)=\infty \text{ \ when }r\leq \sigma _{i},\quad (i=\{1,2\})
	\\
	&\varphi _{ss}(r) =-\frac{c_{ss}}{r^{6}},\text{ where }r>\sigma _{s}\text{ \
		and }\varphi _{ss}(r)=\infty \text{ \ when }r\leq \sigma _{s}\text{,} \\
	&\varphi _{12}(r) =-\frac{c_{12}}{r^{6}},\text{ where }r>\delta _{12}=\frac{%
		\sigma _{1}+\sigma _{2}}{2}\text{ \ and }\varphi _{12}(r)=\infty \text{ \ when }%
	r\leq \delta _{12} \text{,} \\
	&\varphi _{is}(r) =-\frac{c_{is}}{r^{6}},\text{ where }r>\delta _{is}=\frac{%
		\sigma _{i}+\sigma _{s}}{2}\text{ \ and }\varphi _{is}(r)=\infty \text{ \ when }
	r\leq \delta _{is}.
	\end{split}
	\end{align*}
\end{widetext}
The intermolecular coefficients denoted by $c_{11}$, $c_{22}$, $c_{ss}$, $c_{12}$, $c_{1s}$, $c_{2s}$, are associated with Hamaker's constants $A_{ii}$ and $A_{ss}$
defined as 
\begin{equation}
A_{ii}=\pi ^{2}N^{2}c_{ii}\quad \mathrm{and}\quad A_{ss}=\pi ^{2}N^{2}c_{ss} \, ,
\label{Hamaker}
\end{equation}%
where $N\ $is the number of molecules per unit volume \cite{Hamak,Israel}.
An important property of Hamaker constants is that they can be experimentally determined. The Hamaker constant between the two dissimilar
materials can be estimated in term of Hamaker constants of each material. An
  approximation is proposed in \cite{Mitra},
\begin{equation*}
A_{12}=\sqrt{A_{11}A_{22}},\quad A_{1s}=\sqrt{A_{11}A_{ss}},\quad A_{2s}=%
\sqrt{A_{22}A_{ss}}\ .  \label{Hamaker2}
\end{equation*}
Expressions of coefficients $\lambda _{1},\lambda _{2}$ and $\lambda
_{3}$ can be deduced from Eq. \eqref{Lambdas},
\begin{equation*}
{\lambda _{1}=\frac{2}{3}\frac{\pi\, c_{11}}{\sigma _{1}},\quad
\lambda _{2}=\frac{2}{3}\frac{\pi\, c_{22}}{\sigma _{2}},\quad
\lambda _{3}=\frac{4}{3}\frac{\pi\, c_{12}}{\left( \sigma _{1}+\sigma
	_{2}\right)}.  \label{lammda2}}
\end{equation*} 
\noindent In the model of London's forces, coefficients $\gamma _{11},\,
\gamma _{21},\, \gamma _{12},\, \gamma _{22}$ and $\gamma _{32}$ can
also be obtained by using intermolecular coefficients. They are
proposed in \cite{Gouin 2} for solid wall of small curvature.
Nonetheless, the energy behavior \eqref{surfacenergy} being in
the same form for all solid surfaces \cite{degennes}, we will
consider the same expressions for the intermolecular coefficients, 
\begin{equation*}
{\gamma _{11}=\frac{\pi\, c_{1s}}{12\,\delta _{1s}^{2}}\,\rho
_{s},\quad \gamma _{21}=\frac{\pi\, c_{2s}}{12\,\delta _{2s}^{2}}%
\,\rho _{s}\, ,   \label{gamma1}}
\end{equation*}
where $\rho_s$ is the   mass density  of the solid wall,  and
\begin{equation}
{\begin{split}
\gamma _{12}=\frac{\pi\, c_{11}}{12\,\delta _{1s}^{2} }\,,
\quad\gamma _{22}=\frac{\pi\,c_{22}}{12\,\delta _{2s}^{2}} \,,
\\
\gamma _{32}=\frac{\pi\, c_{12}}{24\,\delta _{2s}^{2}}%
\left( \frac{1}{\delta _{1s}^{2}}+\frac{1}{\delta _{2s}^{2}}\right).
\label{gamma2} 
\end{split} }
\end{equation}%

\section{Equations of motions and boundary conditions for inhomogeneous
mixtures of simple fluids}

To obtain the equilibrium equations  of mixtures, it is easy to use a
variational method. The   energy per unit volume is expressed by Eq.
 \eqref{energy capillaire} and the associated energy of domain $\mathcal{%
D}$ is
\begin{equation*}
E_{\mathcal{D}}=\iiint_{\mathcal{D}}\varepsilon \,dv.
\end{equation*}%
The   energy per unit area of the wall is expressed by Eq.
\eqref{surfacenergy} and the associated energy of surface $\mathcal{%
S}$ is
\begin{equation*}
E_{{\mathcal{S}}}=\iint_{{\mathcal{S}}}e\,ds.
\end{equation*}
The  potential of the system {\it fluid mixture--solid wall} is
\begin{equation*}
E=\iiint_{\mathcal{D}}\varepsilon \, dv +\iint_{{\mathcal{S} }}e\, ds.
\end{equation*}%
The total masses of components of an isolated and fixed domain $\mathcal{D}$
are
\begin{equation*}
M_{i}=\int_{\mathcal{D}}\rho _{i}\,dv\quad \mathrm{with}\quad i\in (1,2).
\end{equation*}%
When we neglect external forces, as gravity force, the equilibrium of the
system is reached when the total energy is minimal and we get the  
variational equation
\begin{equation}
\delta E-\mu _{01}\,\delta M_{1}-\mu _{02}\,\delta M_{2}=0, \label{key}
\end{equation}%
where $\mu _{01}$ and $\mu _{02}$ are two constant Lagrange multipliers which
have the physical dimension of chemical potentials.  Domain $ \mathcal{D}$ is the physical domain occupied by the mixture and  used to represent the possible states of a mechanical system of particles  in thermodynamic equilibrium  with a reservoir at given temperature $T$  (we naturally take account of temperature $T$ through term $\varepsilon _{0}(\rho _{1},\rho _{2})$ included in $\varepsilon $). To identify the equilibrium state, the Gibbs free energy is minimized.  Equation \eqref{key} must be
valid for all variations $\delta \rho _{1}$ and $\delta \rho _{2}$  and
consequently,
\begin{equation}
\begin{split}
& \iiint_{\mathcal{D}}\left(  \frac{\partial \varepsilon (\rho _{1},\rho _{2})}{%
\partial \rho _{1}} \delta \rho _{1}+\frac{\partial \varepsilon (\rho
_{1},\rho _{2})}{\partial \rho _{2}}  \delta \rho _{2}- \mu _{01}\delta
\rho _{1}-\mu _{02}\delta \rho _{2}\,\right)\, dv  \\
& +\iint_{\mathcal{S}}\left( \dfrac{\partial e(\rho _{1},\rho _{2}) }{%
\partial \rho _{1}}\,\delta \rho _{1}+\dfrac{\partial e(\rho _{1},\rho _{2})%
}{\partial \rho _{2}}\,\delta \rho _{2}\right) ds=0.  \label{variation} \\
\end{split}%
\end{equation}%
We define
\begin{equation*}
\left\{
\begin{tabular}{ccc}
$\boldsymbol{\phi}_{1}$  $=$  $\lambda _{1}\,\nabla \rho _{1}+\lambda
_{3}\,\nabla \rho _{2},$ \\
&  &  \\
$\boldsymbol{\phi}_{2}$  $=$  $\lambda _{3}\,\nabla \rho _{1}+\lambda
_{2}\,\nabla \rho _{2},$%
\end{tabular}%
\ \right.
\end{equation*}
and
\begin{equation*}
\mu
_{1}(\rho _{1},\rho _{2})=\dfrac{\partial \varepsilon(\rho _{1},\rho _{2})}{\partial \rho _{1}}\equiv\dfrac{%
\partial \varepsilon _{0}(\rho _{1},\rho _{2})}{\partial \rho _{1}}\, ,
\end{equation*}
\begin{equation*}
\mu _{2}(\rho _{1},\rho _{2}) =\dfrac{\partial \varepsilon(\rho _{1},\rho _{2})
}{\partial \rho _{2}}\equiv \dfrac{\partial \varepsilon _{0}(\rho _{1},\rho _{2})}{%
\partial \rho _{2}}\,,
\end{equation*}
which correspond to the  chemical potential of the two fluid
components {at temperature $T$}. Then, Eq. (\ref{variation}) writes
\begin{equation*}
\begin{split}
& \iiint_{\mathcal{D}}\Big[\big(\mu _{1}(\rho _{1},\rho _{2}) -\mu _{01} %
\big)\,\delta \rho _{1}+\boldsymbol{\phi }_{1}\cdot \delta \nabla \rho _{1} \\
&\quad + \big(\mu _{2}(\rho _{1},\rho _{2}) -\mu _{02} \big)\,\delta \rho
_{2}+\boldsymbol{\phi }_{2}\cdot \delta \nabla \rho _{2}\, \Big]\,dv \\
& +\iint_{\mathcal{S}}\left( \dfrac{\partial \ e(\rho _{1},\rho _{2})}{%
\partial \rho _{1}}\,\delta \rho _{1}+\dfrac{\partial e(\rho _{1},\rho _{2})%
}{\partial \rho _{2}}\,\delta \rho _{2}\right) ds=0.
\end{split}%
\end{equation*}
Due to 
\begin{align*}
&\boldsymbol{\phi }_{i}\,\nabla \delta \rho _{i}=\mathrm{{div}} \left(\boldsymbol{\phi }_{i}\,\delta \rho _{i}\right) -\left( \mathrm{div}\boldsymbol{\phi }_{i}\right) \delta \rho
_{i} \ \text{and} \ \nabla \delta \rho _{i} = \delta \nabla\rho _{i}\, ,
\end{align*}
 and using the divergence theorem, we get
\begin{equation*}
\begin{split}
& \sum_{i=1}^{2}\iiint_{\mathcal{D}}\big(\mu _{i}(\rho _{1},\rho _{2})-\mu
_{0i}- {{\rm div}\; {\boldsymbol{\phi }}_{i}\big)\,\delta \rho _{i}\,d\emph{v}} \\
& +\iint_{\mathcal{S}}\left( \dfrac{\partial e(\rho _{1},\rho _{2})}{%
\partial \rho _{i}}+\boldsymbol{n}\cdot \boldsymbol{\phi }_{i}\right) \,\delta \rho
_{i}\,ds=0,
\end{split}%
\end{equation*}%
where $\boldsymbol{n}$ is the external normal to $\mathcal{S}$.
First, we obtain  the equations of equilibrium
\begin{equation*}
\left\{
\begin{tabular}{ccc}
${\rm div}\;{\boldsymbol{\phi}}_{1}$ & $=$ & $\mu _{1}(\rho _{1},\rho _{2})-\mu _{01}$%
, \\
&  &  \\
{\rm div}\;$\boldsymbol{\phi}_{2}$ & $=$ & $\mu _{2}(\rho _{1},\rho _{2})-\mu _{02}$%
, 
\end{tabular}%
\right.
\end{equation*}
\noindent which gives 
\begin{equation}
\left\{
\begin{tabular}{ccc}
$\lambda _{1}\Delta \rho _{1}+\lambda _{3}\Delta \rho _{2}$ & $=$ & $\mu
_{1}(\rho _{1},\rho _{2})-\mu _{01},$ \\
&  &  \\
$\lambda _{3}\Delta \rho _{1}+\lambda _{2}\Delta \rho _{2}$ & $=$ & $\mu
_{2}(\rho _{1},\rho _{2})-\mu _{02},$%
\end{tabular}%
\ \right.  \label{equilibrium}
\end{equation}%
and second, the boundary conditions at surface $\mathcal{S}$,
\begin{equation*}
\frac{\partial e(\rho _{1},\rho _{2})}{\partial \rho _{1}}+ \boldsymbol{n}%
\cdot \boldsymbol{\phi}_{1}=0\quad \mathrm{and}\quad \frac{\partial e(\rho _{1},\rho
_{2})}{\partial \rho _{2}}+\boldsymbol{n}\cdot \boldsymbol{\phi}_{2}=0.
\end{equation*}%
{The previous results --  expressed at equilibrium -- can   be extended to mixture motions by the addition of inertial forces. When we neglect
	the external forces, as gravity force, the two-liquid mixture motions  verify equations extended from the equilibrium case and presented
	in \cite{Gouin8}. They are in the form}
\begin{equation}
\left\{
\begin{array}{c}
{\boldsymbol{a}_{1}}+\nabla \lbrack \,\mu _{1}(\rho _{1})-\lambda
_{1}\,\Delta \rho _{1}\,-\lambda _{3}\,\Delta \rho _{2}\,]=\nu _{1}\,\Delta {%
\boldsymbol{v}_{1}} \label{motion}, \\
\\
{\boldsymbol{a}_{2}}+\nabla \lbrack \,\mu _{2}(\rho _{2})-\lambda
_{3}\,\Delta \rho _{1}\,-\lambda _{2}\,\Delta \rho _{2}\,]=\nu _{2}\,\Delta {%
\boldsymbol{v}_{2}},
\end{array}%
\right.
\end{equation}

\noindent where $\boldsymbol{a}_{i},\ \boldsymbol{v}_{i},\ \nu _{i},\ i\in
\{1,2\}$ denote the accelerations, velocities and   kinematic coefficients
of viscosity of the two-fluid components, respectively. What is unexpected at first, the kinematic
coefficients of viscosity are the same as for bulks \cite{Bocquet}. The
boundary conditions at surface $\mathcal{S} $ are unchanged.\newline
Let us note that chemical potentials naturally introduce  isobaric ensembles.
\section{The chemical potentials in fluid-component bulks}
 
The two-fluid mixture is constituted of simple compressible fluids.
Consequently, pressure and\ chemical potential are the sum of the pressures
and chemical potentials of fluid-components. \newline
In the  fluid component  bulks   corresponding to the plane
liquid-vapor interface at temperature $T$, each chemical potential is denoted by $\mu _{{i0}}$, $i \in \left\{ 1,2\right\} $.
Due to the pressure equation   $p_{i} = p_{i}\,(\rho _{i},T)$ of
fluid-component $i$, it is possible to express $\mu _{{i0}}$ as a function
of $\rho _{i}$.\newline
At given temperature $T$, the associated volume free energy of fluid component $i$, denoted $g_{{i0}}$%
, verifies $g_{{i0}}^{\prime }(\rho _{i})=\mu _{{i0}}(\rho _{i})$.
Potentials $\mu _{i0}$ and $g_{{i0}} $ are defined except an additive
constant and therefore we can choose the conditions
\begin{equation*}
\mu _{{i0}}(\rho _{i\ell })=\mu _{{i0}}(\rho _{iv})=0\quad {\mathrm{and}%
\quad g_{{i0}}(\rho _{i\ell })=g_{{i0}}(\rho _{iv})}=0,
\end{equation*}%
where $\rho _{i\ell }$ and $\rho _{iv}$ are the densities in liquid and
vapor of the bulks corresponding to the plane liquid-vapor interface of fluid-component
$i$. The expressions of thermodynamical potentials $\mu _{{i0}}$ and $%
g_{{i0}}$ can be expanded to the first order near liquid and vapor bulks,
respectively
\begin{equation*}
\mu _{{i0}}(\rho_i )=\frac{C_{i\ell }^{2}}{\rho _{i\ell}}\left( \rho
_{i}-\rho _{i\ell }\right) \quad \mathrm{{and}\quad }\mu _{{i0}}(\rho_i )=%
\frac{C_{iv}^{2}}{\rho _{iv}}\left( \rho _{i}-\rho _{iv}\right) ,
\end{equation*}%
\begin{equation*}
g_{{i0}}(\rho_i )=\frac{C_{i\ell }^{2}}{2\rho _{i\ell }}\left( \rho
_{i}-\rho _{i\ell }\right) ^{2}\quad {\mathrm{and}\quad g_{{i0}}(\rho
_{i})=\frac{C_{iv}^{2}}{2\rho _{iv}}\left( \rho _{i}-\rho _{iv}\right) ^{2}},
\end{equation*}%
where $C_{i\ell }$ and $C_{iv}$ are the isothermal sound-velocities in
liquid and vapor bulks of component $i$ \cite{Gouin 6}. It is
possible to obtain the connection between the liquid bulk of density $\rho
_{i\ell {b}}$ and the vapor bulk of density $\rho _{iv{b}}$ corresponding
to non-planar interfaces (as for spherical bubbles and droplets \cite{Isola2}%
); they are called the mother-bulk densities \cite{Derjaguin}. These equilibria do
not obey the Maxwell rule, but the values of the chemical potential in the
two mother bulks must be equal
\begin{equation*}
\mu _{{i0}}(\rho _{i\ell {b}})=\mu _{{i0}}(\rho _{iv{b}}),  \label{equcp}
\end{equation*}
and determine  the connection between $\rho _{i\ell {b}}$ and
$\rho _{iv {b}}$.\newline
Consequently, we define $\mu _{i\ell {b}}(\rho _{i})$ and $\mu_{iv{b}}(\rho_i )$ as 
\begin{equation*}
\mu _{i\ell {b}}(\rho _{i})=\mu _{_{i0}}(\rho _{i})-\mu _{_{i0}}(\rho
_{i\ell {b}})\equiv \mu _{_{i0}}(\rho _{i})-\mu _{_{i0}}(\rho
_{iv{b}})=\mu _{iv{b}}(\rho _{i})\,.
\end{equation*}
An expansion to the first order near the liquid bulk 
 of density $\rho
_{i\ell {b}}$ and the vapor bulk of density $\rho _{iv{b}}$   yields
\begin{equation}
\mu _{i\ell {b}}(\rho_i)=\frac{C_{i\ell}^{2}}{\rho
_{i\ell b}}\left( \rho _{i}-\rho _{i\ell {b}}\right) \quad
\mathrm{{and} \quad }\mu _{iv b}(\rho _{i})=\frac{C_{iv}^{2}}{\rho
_{iv b}}\left( \rho _{i}-\rho _{iv{b}}\right)  \label{bulk}
\end{equation}%
To the chemical potential $\mu_{i\ell {b}}(\rho _{i})\equiv \mu
_{iv{b}}(\rho _{i})$, we can associate the volume free energies $g_{i\ell
{b}}(\rho _{i})$ and $g_{iv{b}}(\rho _{i})$ which are null for $\rho
_{i\ell {b}}$ and $\rho _{iv{b}}$, respectively,
\begin{equation}
g_{i\ell {b}}(\rho _{i})=g{_{i0}}(\rho _{i})-g{_{i0}}(\rho
_{i\ell {b}})-\mu {_{i0}}(\rho _{i\ell b })(\rho _{i}-\rho
_{i\ell {b}}), \label{freenergy1}
\end{equation}%
\begin{equation}
g_{iv{b}}(\rho _{i})=g_{{i0}}(\rho _{i})-g_{{i0}}(\rho _{iv{b}})-\mu
_{{i0}}(\rho _{iv{b}})(\rho _{i}-\rho _{iv{b}}).  \label{freenergy2}
\end{equation}%
Free energies $g_{i\ell {b}}(\rho _{i})$ and $g_{iv{b}}(\rho _{i})$
are   null for the liquid bulk of density $\rho
_{i\ell {b}}$ and the vapor bulk of density $\rho _{iv{b}}$ of
fluid component $i$, respectively, and contrary to the chemical potentials, they differ by
a constant. Moreover, near the liquid and vapor mother-bulks, the volume
free energies can be expanded as
\begin{equation*}
g_{i\ell {b}}(\rho _{i})=\frac{C_{i\ell }^{2}}{2\rho _{i\ell b }}\left( \rho
_{i}-\rho _{i\ell {b}}\right) ^{2}\ \ \mathrm{{and}\quad  }%
g_{iv{b}}(\rho _{i})=\frac{C_{iv}^{2}}{2\rho _{iv b}}\left( \rho _{i}-\rho
_{iv{b}}\right) ^{2}.
\end{equation*}
To compare the case of a nanotube filled with
liquid-mixture and the case of a  nanotube filled with
vapor-mixture, we chose as \textit{reference  
	volume free energy} of component $i$ the  volume free energy  $g_{i\ell{b}}
(\rho_i)$. Due to $\mu _{{i0}}(\rho _{i\ell b })=\mu
_{{i0}}(\rho _{iv b })$, the difference of Eq. \eqref{freenergy1}
and\, Eq. \eqref{freenergy2} yields
\begin{equation*}
g_{i\ell b}(\rho _{i})-g_{iv{b}}(\rho _{i}) = g_{{i0}}(\rho
_{iv b})-g_{{i0}}(\rho _{i\ell {b}})-\mu _{{i0}}(\rho _{i\ell b
})(\rho _{iv b}-\rho _{i\ell {b}})\,.
\end{equation*}
But, Eq. \eqref{freenergy1} yields
\begin{equation*}
\begin{split}%
& g_{i\ell{b}}(\rho _{i\ell{b}})=0,\\
& g_{i\ell {b}}(\rho _{iv b})=g_{{i0}}(\rho
_{iv b})-g_{{i0}}(\rho _{i\ell {b}})-\mu _{{i0}}(\rho _{i\ell b
})(\rho _{iv b}-\rho _{i\ell {b}}),
\end{split}%
\end{equation*}
and consequently,
\begin{equation}
g_{i\ell{b}}(\rho _{i})=g_{iv{b}}(\rho _{i})+g_{i\ell{b}}(\rho
_{iv{b}}) .
\label{Refrence1}
\end{equation}%
From Eq. \eqref{Refrence1},
the term $g_{i\ell{b}}(\rho _{iv{b}})$ represents the difference
between   the  volume free energies \eqref{freenergy1} and \eqref{freenergy2}.\\

\section{water-ethanol mixture in carbon nanotubes}

We denote by $\mathcal{M}$, $\mathcal{L}$ and $\mathcal{T}$  the mass,  length and   time dimensions,  respectively. The dimensions of other physical quantities are,
 
molecular coefficients: $\mathcal{M}   \mathcal{L}^8
\mathcal{T}^{-2}$,
 
 masses per unit of volume:
$\mathcal{M}   \mathcal{L}^{-3}$,

 isothermal
sound-velocities: $\mathcal{M}   \mathcal{T}^{-1}$. \newline
The values of physical quantities are expressed for water and ethanol   at $20^\circ $ Celsius. These values correspond to molecular coefficients, molecular
diameters, molecular masses,  mass densities, and isothermal sound-velocities of fluid components for liquids and vapors.  They are obtained from the books of Israelaschvili  \cite{Israel}, the {\it Handbook of Chemistry and Physics} \cite{Handbook}, and the {\it Microfluidics and Nanophysics Handbook} \cite{Mitra}, and are indicated in Table I. Carbon values are referred to
 nanotubes from Matsumoto
 \emph{et al}'s paper \cite{Matsu}.  Molecular coefficients are deduced from the Hamaker  constants thanks to relations \eqref{Hamaker}.  The values of Table I together with Eqs. 
\eqref{Hamaker} to
\eqref{gamma2} allow us to obtain the values of $\lambda_1, \lambda_2,
\lambda_3, \gamma _{11},\gamma _{21},\gamma _{12},\gamma _{22},\gamma _{32}$ and those given in  Table \ref{TableKey2}.
\small
\begin{widetext}
\begin{table}
\begin{center}
	\centering{
\begin{tabular}{|c|c|c|c|c|c|}
\hline
\multicolumn{1}{|c|}{\scriptsize Physical constants} &  {\scriptsize{Intermolecular coefficient}} & 
 {\scriptsize{molecular diameter}}  &{ \scriptsize{molecular mass}} & {\scriptsize{Bulk density}} & {\scriptsize{Isothermal sound velocity}}  \\ \hline
\multicolumn{1}{|c|}{\scriptsize Liquid-water} & $1.4\times 10^{-77}$ & $%
2.8\times 10^{-10}$ & $2.99\times 10^{-26}$ & $998$ &{$%
1.478\times 10^{3}$} \\ \hline
\multicolumn{1}{|c|}{\scriptsize Vapor-water} & $1.4\times 10^{-77}$ & $%
2.8\times 10^{-10}$ & $2.99\times 10^{-26}$ & $9.7 \times 10^{-3}$ &
{$3.70\times 10^{2}$} \\ \hline
\multicolumn{1}{|c|}{\scriptsize Liquid-ethanol} & $3.71\times 10^{-77}$ & $%
4.69\times 10^{-10}$ & $7.64\times 10^{-26}$ & $789$ &{$%
1.162\times 10^{3}$} \\ \hline
\multicolumn{1}{|c|}{\scriptsize Vapor-ethanol} & $3.71\times 10^{-77}$ & $%
4.69\times 10^{-10}$ & $7.64\times 10^{-26}$ & $1.09 \times 10^{-1}$ &
{$2.30\times 10^{2}$} \\ \hline
\multicolumn{1}{|c|}{\scriptsize Nanotube-carbon} & $0.1014\times 10^{-77}$
& $1.30\times 10^{-10}$ & $1.99\times 10^{-26}$ & $2000$ &
{$--- $} \\ \hline
\end{tabular}
% \vskip 0.1cm
\caption{The physical values associated with water, ethanol and carbon are
expressed in {\emph{M.K.S. units}} (meter, kilogram, second). {In the unit system, we notice
the very small values of intermolecular coefficients, molecular diameters and molecular masses.}}}
\label{TableKey1}
\end{center}
\end{table}
\end{widetext}
\normalsize
\begin{table}
\centering{
\begin{tabular}{|c|c|c|c|c|c}
\hline
\multicolumn{1}{|c|}{\scriptsize Physical constants} & $\lambda_{1}$ & $%
\lambda_{2}$ & $\lambda_{3}$ & {$\gamma_{11}$}   \\
\hline
\multicolumn{1}{|c|}{\scriptsize Numerical values} & $1.17\times 10^{-16}$ &
$0.284\times 10^{-16}$ & $0.49\times 10^{-16}$ & $77\times 10^{-6}$    \\
\hline
\multicolumn{1}{|c|}{\scriptsize Physical constants} & $\gamma_{21}$ & $%
\gamma_{12}$ & $\gamma_{22}$ & {$\gamma_{32}$}    \\
\hline
\multicolumn{1}{|c|}{\scriptsize Numerical values} & $22\times 10^{-6}$ & $%
9.7\times 10^{-8}$ & $1.8\times 10^{-8}$ &{$4.0\times
10^{-8}$}    \\ \hline
\end{tabular}
\vskip 0.1cm }
\caption{ The physical values associated with the contact of water, ethanol
and carbon expressed in {\emph{M.K.S. units}} (meter, kilogram, second).}
\label{TableKey2}
\end{table}

\subsection{\label{5A}{Liquid-mixture at equilibrium in carbon nanotubes}}

In this  section, we compare  -- at equilibrium --  the total free energy of a carbon nanotube filled with a mixture of water and ethanol in liquid, vapor and liquid-vapor conditions. The   reference volume free energy  is $g_{i\ell{b}}(\rho _{i})$. 
Including the wall-energy, the total free energy per unit of
length of the nanotube can be approximated by considering densities $\rho_{1}$ (for water) and $\rho_{2}$ (for ethanol) of the two components closely equal to $\rho_{1\ell b }$ and $\rho_{2\ell b }$ or  $\rho_{1v b }$ and $\rho_{2v b}$    for the liquid and the vapor phases, respectively.\\

First, we compare the case of a nanotube filled with liquid mixture and a nanotube filled with vapor mixture. \\

\noindent (i) \quad In the case of a nanotube filled of liquid-mixture, total free energy $E_1$ per unit of length is,
\begin{align*}
\begin{split}
& E_{1} \approx\Pi \left( \rho
_{\ell{b}}\right)\pi \, R^{2}\left[ g_{1\ell{b}}(\rho _{1\ell{b}})+g_{2\ell{b}}(\rho
_{2\ell{b}})\right]  
\\
& + 2\ \pi \ R\ \Big[ -\gamma _{11}\,\rho _{1\ell{b}}-\gamma _{21}\,\rho _{2\ell{b}} 
\\%
&+ \frac{1}{2}\left( \gamma _{12}\,\rho _{1\ell{b}}^{2}+\gamma _{22}\,\rho
_{2\ell{b}}^{2}+
2\gamma _{32}\,\rho _{1\ell{b}}\,\rho _{2\ell{b}}\right) \Big],
\end{split}
\end{align*}
where $g_{1\ell{b}}(\rho _{1\ell{b}})=0$\, and\, $g_{2\ell{b}}(\rho _{2\ell{b}})=0$.\newline

\noindent (ii)\quad In the case of a nanotube filled of
the vapor-mixture, total free energy $E_2$ per unit of length is,
\begin{align*}
\begin{split}
& E_{2}\approx\pi \ R^{2}\left[ g_{1\ell{b}}(\rho _{1v{b}})+g_{2\ell{b}}(\rho
_{2v{b}})\right]  
\\
&+ 2\ \pi \ R\ \Big[ -\gamma _{11}\rho _{1v{b}}-\gamma _{21}\rho _{2v{b}}
\\
&+%
\frac{1}{2}\left( \gamma _{12}\,\rho _{1v{b}}^{2}+\gamma _{22}\,\rho
_{2v{b}}^{2}+2\gamma _{32}\,\rho _{1v{b}}\,\rho _{2v{b}}\right)
\Big].
\end{split}
\end{align*}%
Due to the densities of vapor components,
\begin{align*}
\begin{split}
 -\gamma _{11}\,\rho _{1v{b}}-\gamma _{21}\,\rho _{1v{b}}
 +\frac{1}{2}\left( \gamma _{12}\,\rho _{1v{b}}^{2}+\gamma _{22}\,\rho
_{2v{b}}^{2}+2\gamma _{32}\,\rho _{1v{b}}\,\rho _{2v{b}}\right)  
 \end{split}
\end{align*}%
is negligible in the estimation of energy {with respect to the first two terms}. From the expression of   partial pressure $p_i$ of component $i$, $i\in \{1, 2\}$,
\begin{equation*}
p_i(\rho_i)=\rho_i\, \mu_{{i0}}(\rho_i)-g_{i0}(\rho_i)+p_{i0}
\end{equation*}
where $p_{i0}$ is the common value of the pressure in liquid and vapor bulks of plane interface. Equation \eqref{freenergy1} yields,
\begin{equation*}
\begin{split}
g_{1\ell{b}}(\rho _{1v{b}})+g_{2\ell{b}}(\rho _{2v{b}})=& \ p_{1}\left(
\rho _{1\ell{b}}\right) -p_{1}\left( \rho _{1v{b}}\right)  \\  +& \  
p_{2}\left( \rho _{2\ell{b}}\right) -p_{2}\left( \rho _{2v{b}}\right) .
\end{split}
\end{equation*}%
We denote ${\it \Pi}_b$,
\begin{eqnarray*}
 && {\it \Pi}_b   =   p\left( \rho _{v{b}}\right) -p\left( \rho
_{\ell{b}}\right)= \\
&& p_{1}\left( \rho _{1v{b}}\right)   -   p_{1}\left( \rho _{1\ell{b}}\right) %
 +   p_{2}\left( \rho _{2v{b}}\right) -p_{2}\left( \rho
_{2\ell{b}}\right)  ,
\end{eqnarray*}%
the difference between   pressure $p\left( \rho _{v{b}}\right)$ of the vapor mixture and pressure $p\left( \rho _{\ell{b}}\right)$ of the 
liquid mixture.
Pressure ${\it \Pi}_b$ is 
called the mixture \textit{disjoining-pressure} \cite{Derjaguin,Gouin 6}.
Consequently,
\begin{align*}
& E_{2}-E_{1}  \approx
2\ \pi \ R\ \Big[ \gamma _{11}\,\rho
_{1\ell{b}}+\gamma _{21}\,\rho _{2\ell{b}}\\
&-\frac{1}{2}\left( \gamma
_{12}\,\rho _{1\ell{b}}^{2}+\gamma
_{22}\,\rho _{2\ell{b}}^{2}+2\,\gamma _{32}\,\rho _{1\ell{b}}\,\rho _{2\ell{b}}\right) %
\Big]   
  -\pi \ R^{2\ }{\it \Pi}_b \,   ,
\end{align*}%
and the vapor mixture is more energetic than the liquid mixture if
\begin{align}
 E_{2}>E_{1}\quad \Longleftrightarrow\quad
{\it \Pi}_b  < \frac{1}{R}\ \Big[2\, \gamma _{11}\,\rho
_{1\ell{b}}+2\,\gamma _{21}\,\rho _{2\ell{b}}\notag \\
- \left( \gamma
_{12}\,\rho _{1\ell{b}}^{2}+\gamma
_{22}\,\rho _{2\ell{b}}^{2}+2\,\gamma _{32}\,\rho _{1\ell{b}}\,\rho _{2\ell{b}}\right) %
\Big] .  \label{inequality}
\end{align}%
The ratio between volume proportion of water and ethanol is denoted $c$.
In the bulks, the water density is\ \, $\rho_{1\ell b }=c\,\rho_{1\ell}$ and the ethanol
density is\, $\rho_{2\ell {b}}=(1-c)\,\rho_{2\ell}$. By taking account of Table \ref{TableKey2},
we  calculated the most unfavorable case for inequality \eqref{inequality} to be verified. One obtain $R_{0}=50$ nm (for a microtube of $0.1$ micron diameter) and we get 
\begin{equation*}
{\it \Pi}_b \equiv  p\left( \rho
_{v{b}}\right) -p\left( \rho _{\ell{b}}\right) \approx 5 \times 10^{5}\, {\rm Pascal}   
=5\,  {\rm atmospheres}. \label{Cond1}
\end{equation*} 
This result comes from the   wall quality of carbon nanotubes: terms $\gamma _{11} $\, and $%
\gamma _{21}$ take advantage of terms associated with $\gamma
_{12},\ \gamma _{22}$ and $\gamma _{32}$, and  consequently the energy of vapor  mixture is greater than the
energy of liquid mixture. 

\begin{figure}
%\flushright
\includegraphics[width=6.5 cm]{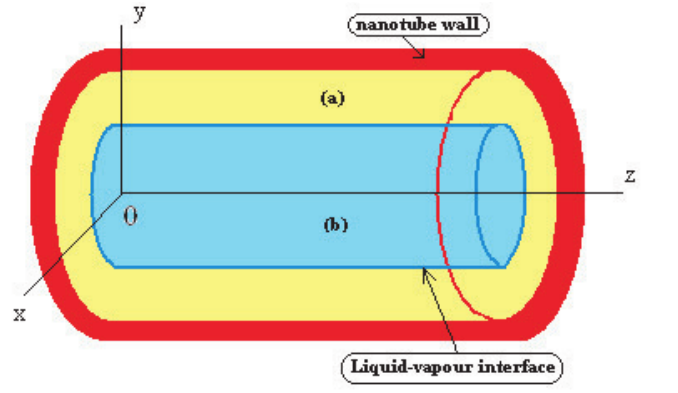}
\caption{Two-phase fluid-component in a
nanotube: The nanotube is simultaneously filled with two phases
liquid and vapor of  fluid-components. The two phases
(a) and (b) are separated by a { cylindrical interface}.} \label{Fig.
1}
\end{figure}
Second, we compare the case of a nanotube filled with 
  liquid mixture and a nanotube filled with a liquid-vapor mixture.
When a two-phase mixture is considered, we assume that an
interface appears between the liquid and vapor phases. { To estimate the value of the liquid-vapor interface, we consider the smallest possible interface area. The
smallest area of the liquid-vapor interface separating the two
phases corresponds to a material surface represented by a cylinder
with the same axis as the nanotube axis}. The interface has a positive
surface energy $\gamma $ increasing the total fluid energy inside the nanotube (see Fig. \ref{Fig. 1})\\

(i) \quad  When domain  (a) is liquid and domain
(b) is vapor, the energy of
the wall is approximatively the same as for liquid mixture and the difference of total free energies between liquid mixture and liquid-vapor mixture is approximately
\begin{equation*}
E_{3}-E_{1}\approx\pi \, r_{1}\left[\, 2\, \gamma - r_{1}\, {\it \Pi}_b\right] ,
\end{equation*}
where $E_{3}$ is the energy of the two-phase mixture and $r_{1}$
is the radius of domain (b).

\noindent (ii) \quad  When domain (a) is vapor and
domain (b) is liquid, the difference of total free
energies is approximately%
\begin{align*}
& E_{4}-E_{1} \approx 2\, \pi \, r_{1}\,  \gamma - \pi \, \left(
R^{2}-r_{1}^{2}\right) {\it \Pi}_b\\
&+ 2\, \pi \, R\,  \Big[\, \gamma _{11}\rho _{1v{b}}+\gamma _{21}\rho _{2v{b}}\\
& -
\frac{1}{2}\left( \gamma _{12}\rho _{1v{b}}^{2}+\gamma _{22}\rho
_{2v{b}}^{2}+2\gamma _{32}\rho _{1v{b}}\rho _{2v{b}}\right) \Big]\,,
\end{align*}
where $E_{4}$ is the energy of the two-phase mixture.
The positive interfacial energy associated with $\gamma$ increases the total energy of the liquid-vapor mixture. \newline
 In all   cases, when $\gamma>12.5 \times 10^{-3}$  N $\times$ m$^{-1}$ (corresponding to a value significantly lower than the surface tension of any water-ethanol mixture at $20^\circ $ C \cite{Handbook}), and condition \eqref{inequality} is verified, we get
 \begin{equation*}
 E_3 > E_1\quad {\rm and}\quad E_4 > E_1\,.
 \end{equation*}
In the following, we consider only carbon nanotubes such that   previous conditions are verified. Consequently  { - at equilibrium - } they are filled with liquid mixture
of water and ethanol.

\subsection{{Numerical calculations at equilibrium}}

We consider  cylindrical coordinates $(r,\theta ,z)$\,  where $z$
denotes the coordinate of the nanotube axis. Due to the symmetry of
revolution and far from  nanotube extremities, the densities depend only
of $z$ and in the case of liquid mixture, the equations of equilibrium \eqref{equilibrium} can be written 
 by taking  account of Eq. \eqref{bulk},
\begin{equation}
\left\{
\begin{tabular}{ccc}
$\displaystyle\lambda _{1}\left( \frac{d^{2}\rho _{1}}{dr^{2}}+\frac{1}{r}%
\frac{d\rho _{1}}{dr}\right) +\lambda _{3}\left( \frac{d^{2}\rho _{2}}{dr^{2}%
}+\frac{1}{r}\frac{d\rho _{2}}{dr}\right) $ & $=$ & $\displaystyle\frac{%
c_{1\ell }^{2}}{\rho _{1\ell b }}\left( \rho _{1}-\rho _{1\ell {b}}\right) ,$
\\
&  &  \\
$\displaystyle\lambda _{3}\left( \frac{d^{2}\rho _{1}}{dr^{2}}+\frac{1}{r}%
\frac{d\rho _{1}}{dr}\right) +\lambda _{2}\left( \frac{d^{2}\rho _{2}}{dr^{2}%
}+\frac{1}{r}\frac{d\rho _{2}}{dr}\right) $ & $=$ & $\displaystyle\frac{%
c_{2\ell }^{2}}{\rho _{2\ell b }}\left( \rho _{2}-\rho _{2\ell {b}}\right) ,$%
\end{tabular}%
\ \right.  \label{equ1}
\end{equation}
where $\rho_{1}$ and $\rho_{2}$ denote the densities associated with water and ethanol, respectively.
The symmetry of densities at the $z$-axis implies,
\begin{equation}
\mathrm{At}\ \ r=0,\quad \frac{d\rho _{1}}{dr}=0\quad
\mathrm{and}\quad \frac{d\rho _{1}}{dr}=0 \label{BC0}
\end{equation}%
and the boundary conditions at the nanotubes wall --   where $R$ is the nanotube radius -- yield,
\begin{equation}
\mathrm{At}\ \ r=R,\quad \left\{
\begin{tabular}{ccc}
$\displaystyle\lambda _{1}\frac{d\rho _{1}}{dr}+\lambda _{3}\frac{d\rho _{2}%
}{dr}$ & $=$ & $\gamma _{11}-\gamma _{12}\,\rho _{1}-\gamma _{32}\,\rho _{2},$
\\
&  &  \\
$\displaystyle\lambda _{3}\frac{d\rho _{1}}{dr}+\lambda _{2}\frac{d\rho _{2}%
}{dr}$ & $=$ & $\gamma _{21}-\gamma _{32}\,\rho _{1}-\gamma _{22}\,\rho _{2}.$%
\end{tabular}%
\ \right.  \label{BC1}
\end{equation}
\noindent For nanotubes, the values obtained in Table
\ref{TableKey2} are not expressed in convenient units. To obtain
units adapted to the numerical computations, we consider an unit
system such that  the   mass density, the
isothermal sound-velocity of liquid-water and $\lambda_1$ are
equal to 1. Table \ref{TableKey3} indicates the corresponding
values of units of length, mass and time. We {name} these units,
\emph{molecular capillary units} at temperature  {$T=20^\circ $ C}.
\noindent We notice that the units of length mass and time are of the same   order
as molecular diameters, masses of molecules and times associated with the
mean free paths. In molecular capillary units,
the values of $\lambda_1, \lambda_2, \lambda_3, \gamma _{11},\gamma
_{21},\gamma _{12},\gamma _{22}$,and $\gamma _{32}$ are indicated in Table \ref%
{TableKey4}.
\begin{table}
\centering{ $%
\begin{tabular}{|c|c|c|c|}
\hline
\multicolumn{1}{|c|}{\scriptsize Physical units} & \scriptsize  unit of length & \scriptsize  unit of
mass & \scriptsize  unit of time \\ \hline
\multicolumn{1}{|c|}{\scriptsize Numerical values} & $2.31\times 10^{-10}$ m
& $1.23\times 10^{-26}$ Kg & $1.56\times 10^{-13}$ s \\ \hline
\end{tabular}
$ \vskip 0.1cm }
\caption{ The new {\emph{molecular capillary units}}, at temperature  {$T=20^\circ $ C}, associated with
liquid-water are deduced from the Table I and
expressed in {\emph{M.K.S. units}} { (meter, kilogram and second)}.}
\label{TableKey3}
\end{table}

\begin{table}[h]
\centering{\footnotesize $%
\begin{tabular}{|c|c|c|c|c|c|c|c|c|c|c|}
\hline
\multicolumn{1}{|c|}{\scriptsize Physical constants} & $\lambda_{1}$ & $%
\lambda_{2}$ & $\lambda_{3}$ & $\gamma_{11}$ & $\gamma_{21}$ &
$\gamma_{12}$ & $\gamma_{22}$ &
$\gamma_{32}$   \\ \hline
\multicolumn{1}{|c|}{\scriptsize Numerical values} & $1$ & $0.24$
& $0.42$ & $146$ & $42$ & $3.2\times 10^{-3}$ & $6\times 10^{-4}$ &
$1.3\times 10^{-3}$  \\ \hline
\end{tabular}
$ \vskip 0.1cm }
\caption{ The physical values associated with water, ethanol and carbon are
expressed in {\emph{molecular capillarity units}}. Moreover, the isothermal
sound velocities are   $C_{{\mathrm{liquid-water}}} =1$ and $C_{{\mathrm{%
liquid-ethanol}}} =0.786$.}
\label{TableKey4}
\end{table}
\noindent To  easily compare the profiles of densities, we have normalized
	the diameter with respect to  radius $R$ and therefore for  all figures, the $x$-axis is drawn between 0 and 1 whatever is the  real value of the nanotube diameter. 

	\begin{table}[h]
		\centering{$%
			\begin{tabular}{|c|c|c|c|c|c|c|c|c|c|c|}
			\hline \multicolumn{1}{|c|}{\scriptsize $c\,\downarrow\,
				2\,R\,\rightarrow$} &  1\,nm  &  2\,nm &  3\,nm &  5\,nm &
			10\,nm &  20\,nm &  50\,nm &
			{ 100\,nm}   \\ \hline
			\multicolumn{1}{|c|}{\scriptsize $c=0.3$} & 2.7\% & 1.4\% & 1\% &
			0.6\%
			& 0.3\% &0.15\% & 0.1\% &{0.05\%}  \\
			\hline \multicolumn{1}{|c|}{\scriptsize $c=0.5$} & 5\% & 2.5\% &
			1.7\%
			& 1\% & 0.5\% & 0.3\% & 0.2\% &{0.1\%}  \\
			\hline \multicolumn{1}{|c|}{\scriptsize $c=0.7$} & 8.5\% & 4.4\% &
			3\%
			& 1.8\% & 1\% & 0.5\% &  0.37\% &{0.27\%}  \\
			\hline
			\end{tabular}
			$ \vskip 0.1cm } \caption{Percentage of the increase of volume
			ethanol with respect to volume water for different nanotube
			diameters and {values of   ratio  $c$ between volume proportion of water and ethanol}. } \label{TableKeyV}
	\end{table}
\noindent We calculate the profile of densities of the liquid
mixture composed of water and ethanol   wetting carbon nanotubes. As in Section \ref{5A}, the ratio
between volume proportion of water and ethanol is denoted $c$,
in the bulks, the water density is\ \, $\rho_{1\ell b }=c\,\rho_{1\ell}$ and the ethanol
density is\, $\rho_{2\ell {b}}=(1-c)\,\rho_{2\ell}$. The profiles of densities are given by equilibrium equations \eqref{equ1}
and boundary conditions    
\eqref{BC0}--\eqref{BC1}.
\begin{figure}
	\begin{center}
		\includegraphics[width=8cm]{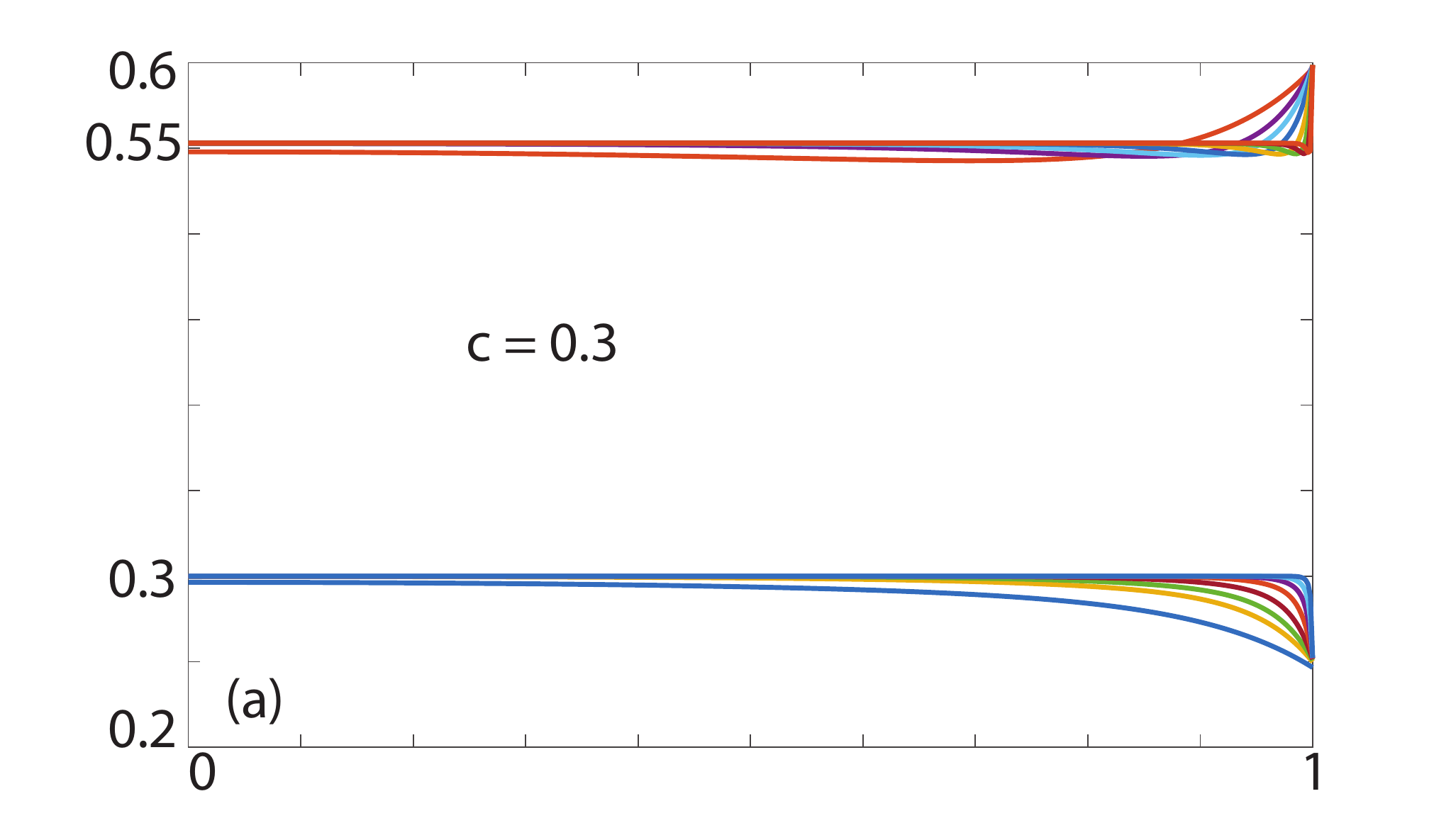}
			\includegraphics[width=8cm]{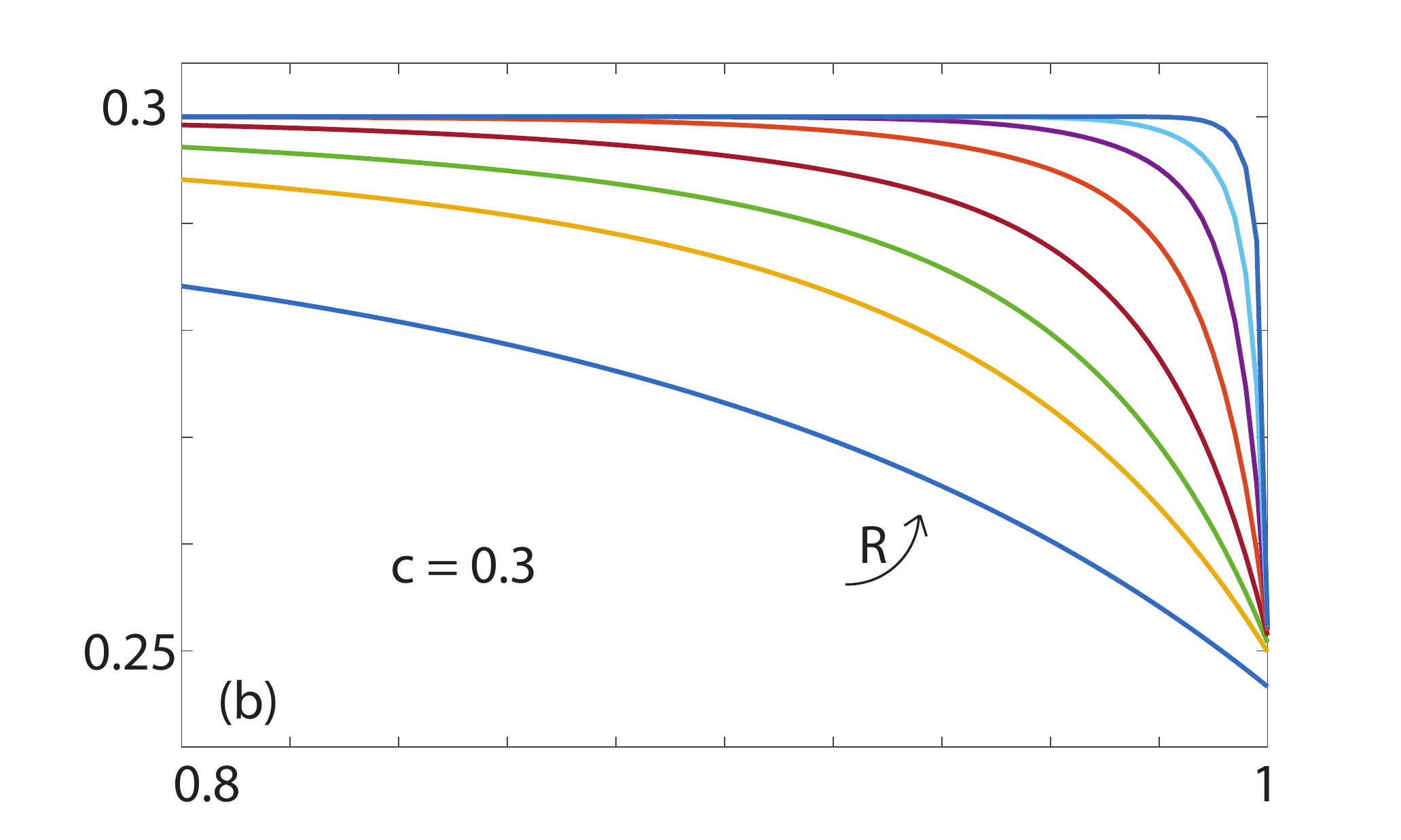}
		\includegraphics[width=8cm]{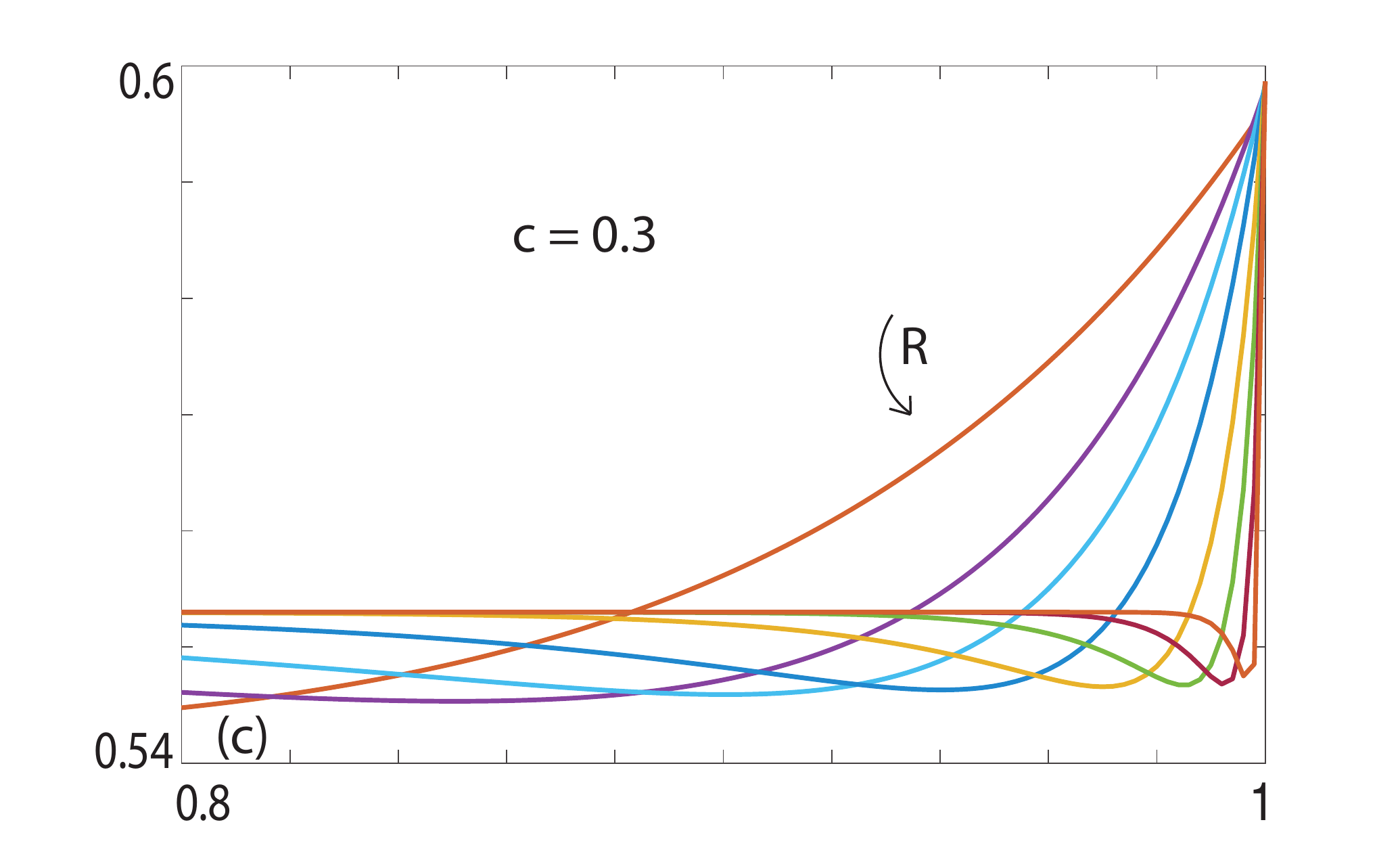}
	\end{center}
	\caption{\footnotesize    {\bf Case $c = 0.3$ (corresponding to the volume proportion between water and ethanol  in the mixture bulk outside the carbon nanotube).} \\  {The different graphs represent the change of the ratio between volume proportion of water and ethanol  versus r / R (representing the normalized distance to the carbon wall), when the radius nanotube is successively  equal to $R = 0.5$ at $50$ nm. {\it We recall that - to draw all the diameter cases on the same figure -  we have normalized
		the diameter with respect to  radius $R$ and therefore, the $x$-axis is drawn between 0 and 1 for any  real value of the nanotube diameter}.}\\ 
	 {The  graph  (a)  shows the density profiles between the nanotube axis and the carbon wall ($r/R \in [0,1]$). The second graph (b) enlarges the water density   near the wall ($r/R \in [0.8,1]$). The third graph (c) enlarges the ethanol density near the wall ($r/R \in [0.8,1]$).   In the second and third graphs, the curved arrows indicate the different graphs for increasing values of R and for the eight nanotube radius-values corresponding to $2R \in \{1, 2, 3, 5, 10, 20, 50,100\}$ in nanometers.}}\label{Fig. 2a}
\end{figure}
\begin{figure}
	\begin{center}
		\includegraphics[width=8cm]{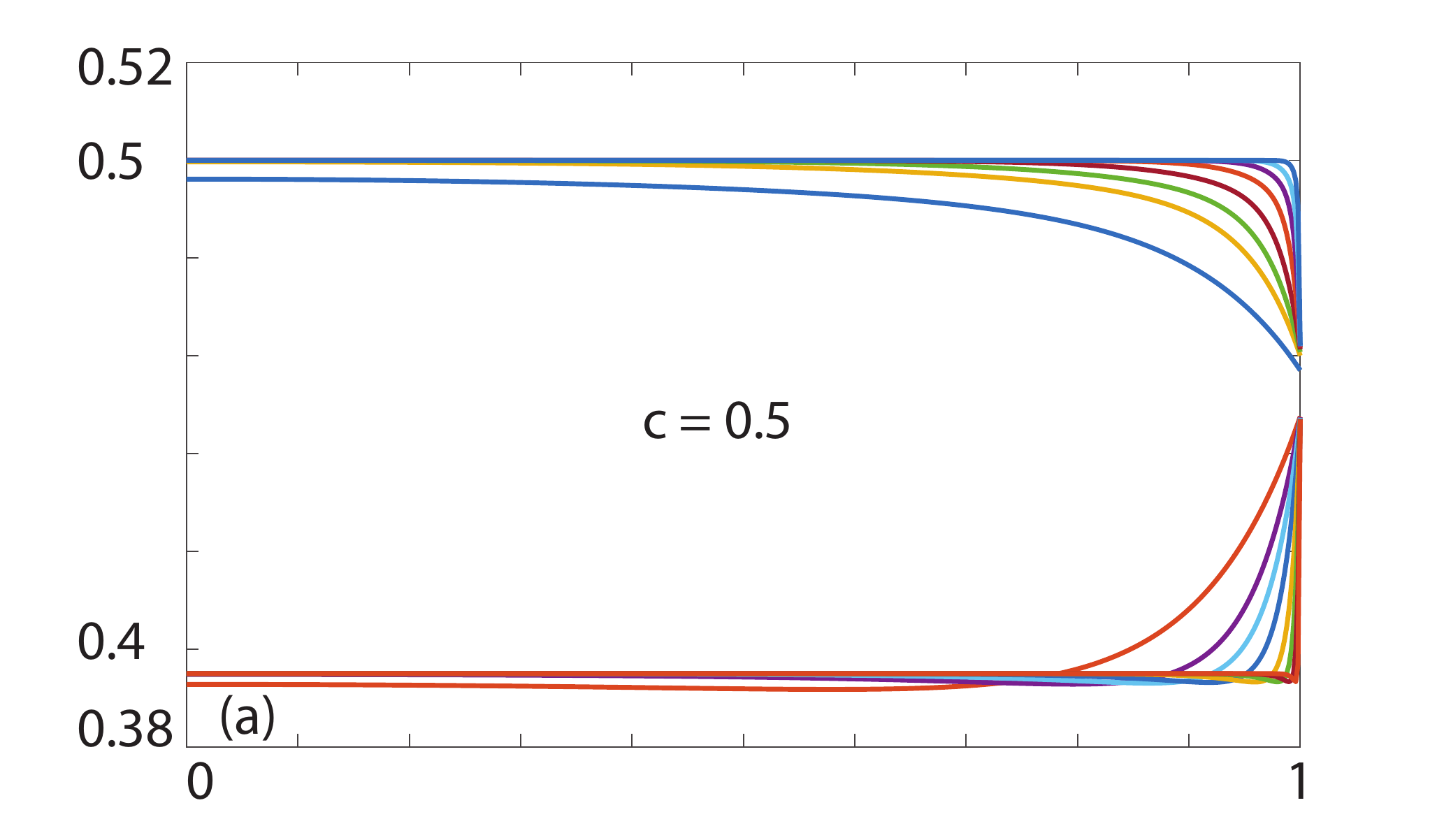} 
		\includegraphics[width=7cm]{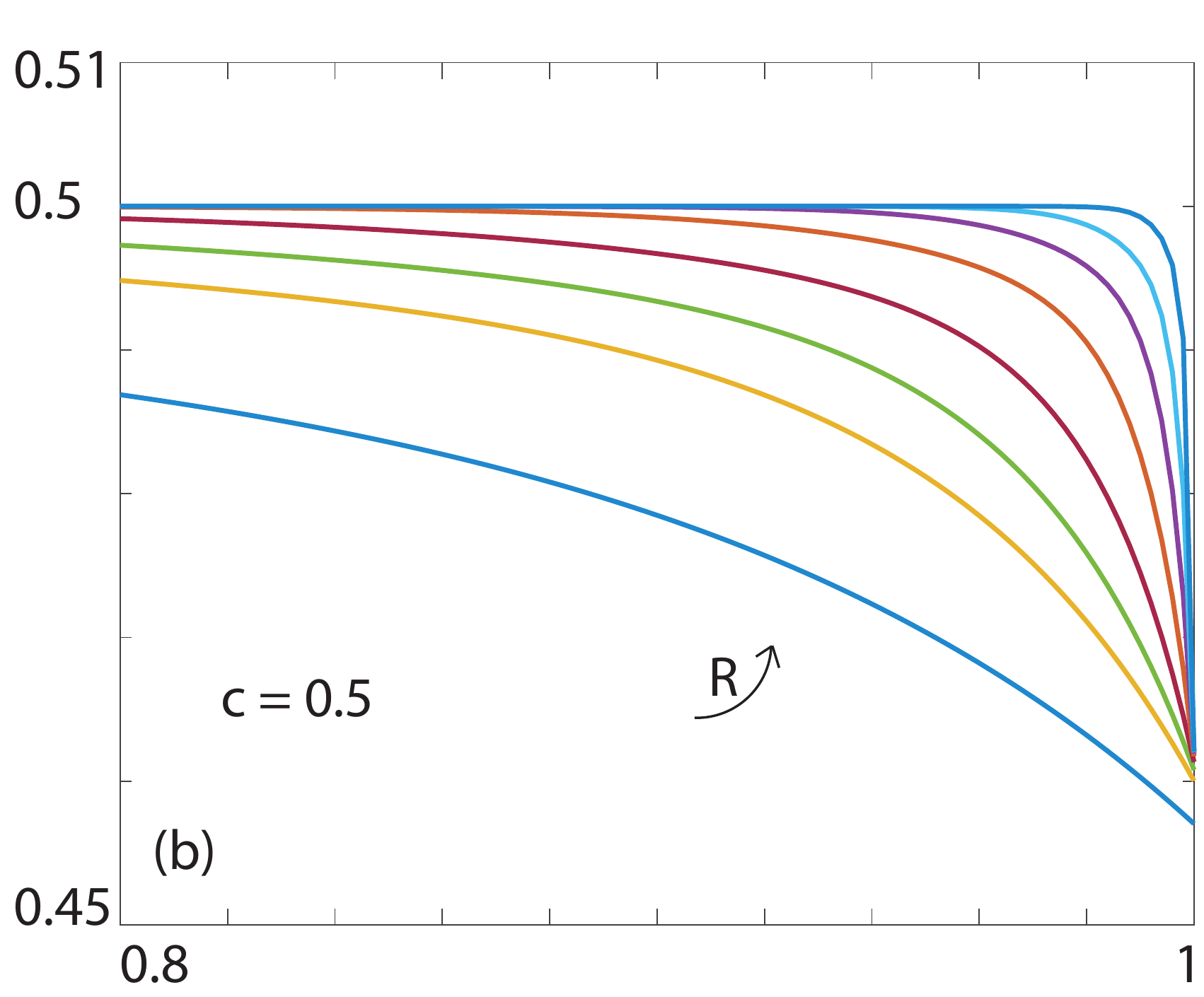} \includegraphics[width=7cm]{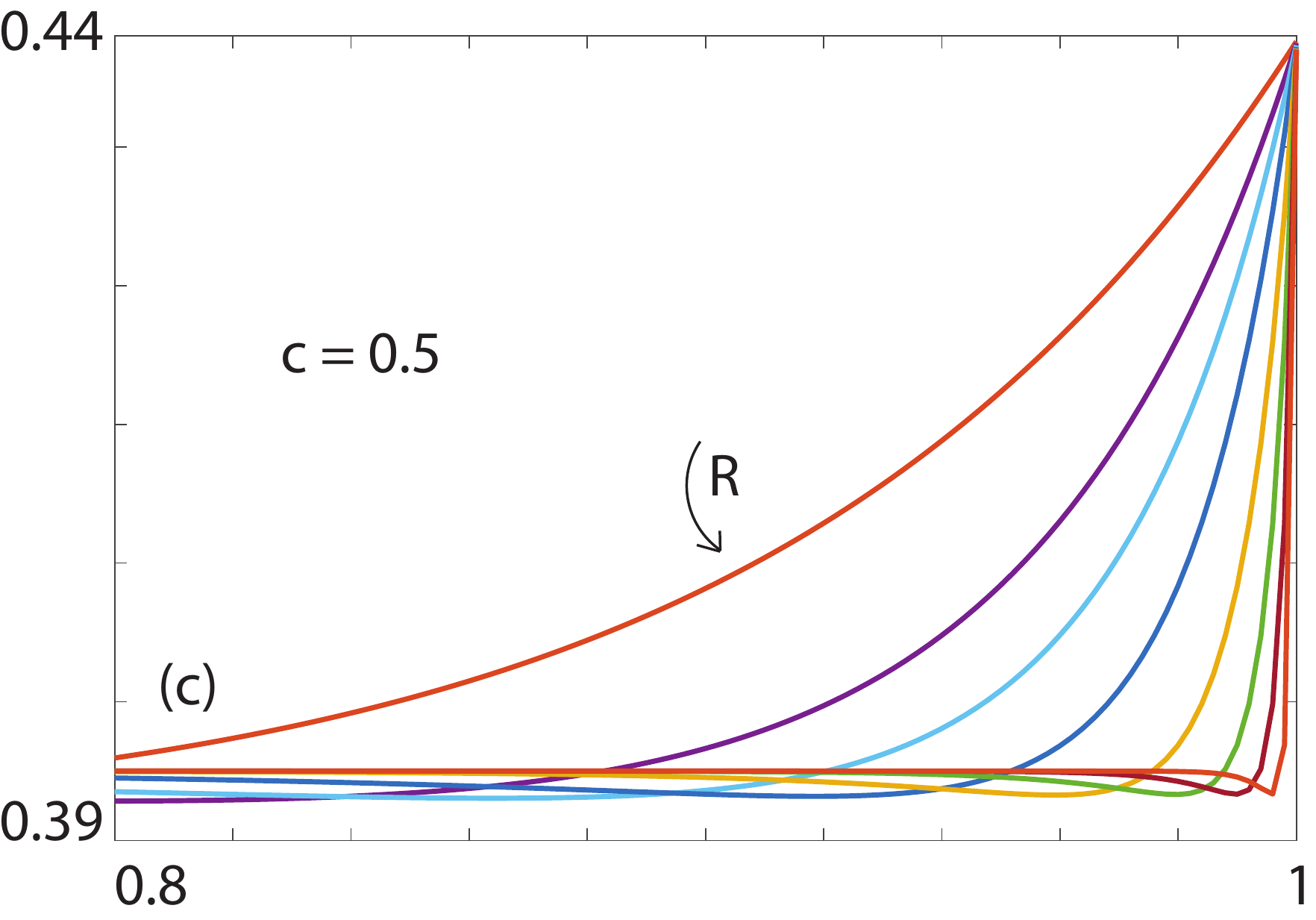}
	\end{center}
	\caption{\footnotesize  {\bf Case $c = 0.5$ (corresponding to the volume proportion between water and ethanol  in the mixture bulk outside the carbon nanotube).}\\  	
	   {Comments are similar to  those  for the caption to Fig. 2}.} \label{Fig. 2b}  
\end{figure}
We calculate three cases of liquid mixtures' composition corresponding {in the bulks} to
$c=0.3,\, 0.5$ and\;    0.7. The density curves depend on   nanotube
diameters. For each value of $c$, we calculate the density
profiles for water and ethanol when the diameters are 1, 2,
3, 5 , 10, 20, 40 and 100 nanometer.  Figures  \ref{Fig. 2a}, \ref{Fig. 2b} and \ref{Fig. 2c}\;
 represent  the densities versus $r/R$ for increasing value of $R$  with  relative up and down magnifications.
\newline
In all   cases, we note that the water density   decreases near the
wall and the profile is monotonic for   $c=0.3$ and $c=0.5$,
and oscillating for   $c=0.7$. For ethanol the density
increases near the wall and the profile is  oscillating for  
$c=0.3$ and $c=0.5$ and monotonic for   $c=0.7$.
\\
Following ratio {$c$ between volume proportion of water and ethanol}  and the nanotube diameters, volume
concentration of ethanol with respect to volume concentration of water
increases. The results are indicated on Table \ref{TableKeyV}. The
effect closely negligible when
$c=0.3$, is noticeable for   small nanotubes  when $c=0.7$,  and
disappears for large nanotubes.

\section{Viscous motions in a nanotube}

Fluid flows through structures like carbon nanotubes must be different from
flows through microscopic and macroscopic structures since, for the latter
flow, the degrees of freedom of fluid molecules can be safely ignored and
the flow in such structures can be characterized by viscosity, density and
other bulk properties. Furthermore, for large-scale systems, the no-slip
boundary condition is often implemented, because the fluid
velocity is negligibly small at the fluid/wall boundary. Reducing the length
scales introduces new phenomena, when the mutual interaction between walls
and fluids must be taken into account \cite{Rafii}. 
 \begin{figure}
 %\begin{center}
  \includegraphics[width=7cm]{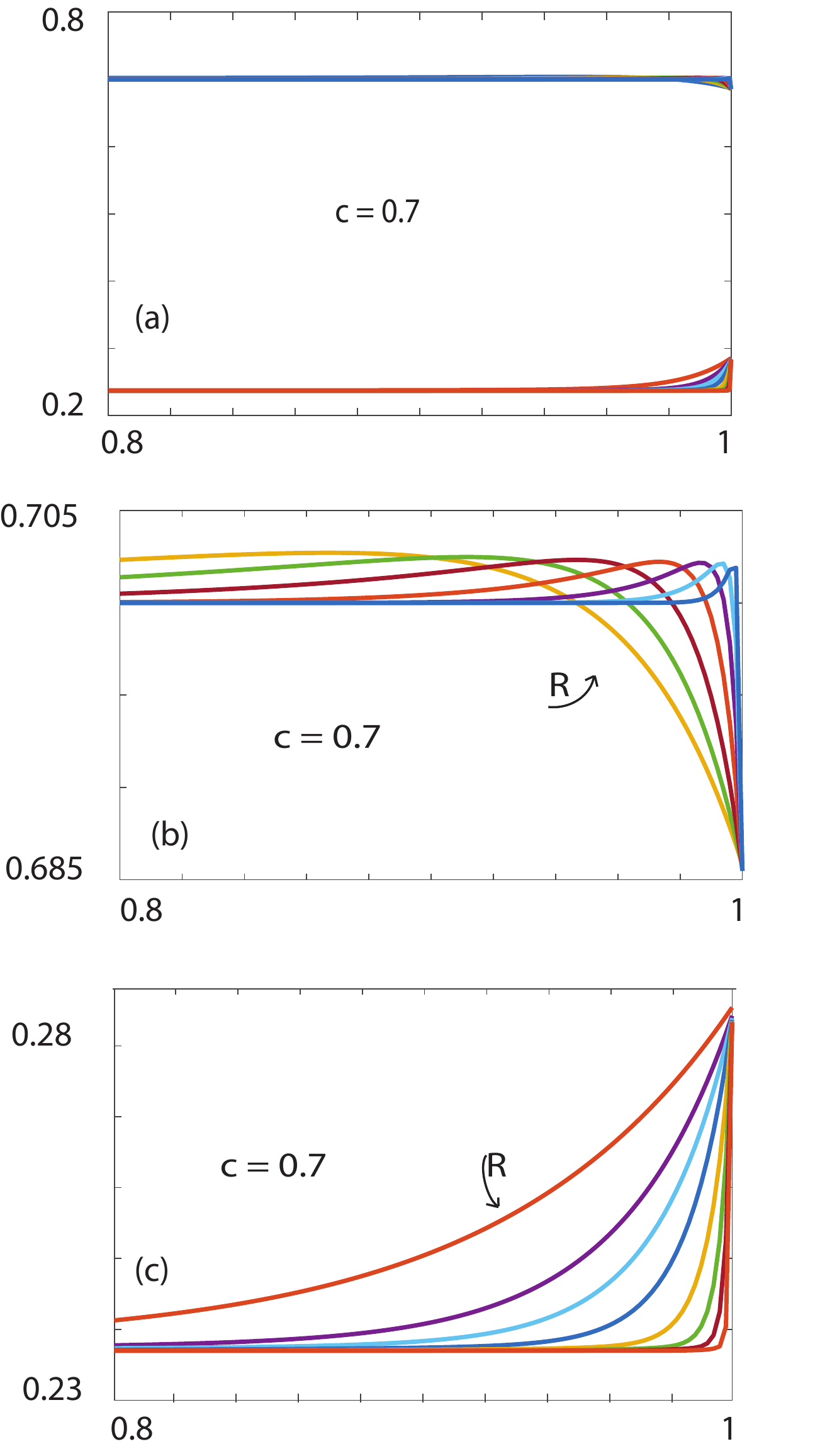} 
%\end{center}
	\caption{\footnotesize   {\bf Case $c = 0.7$ (corresponding to the volume proportion between water and ethanol  in the mixture bulk outside the carbon nanotube).}\\  	
	  {Comments are similar to  those  for the caption to Fig. 2}.} \label{Fig. 2c}
\end{figure}
We consider the permanent and laminar motions of viscous capillary liquids
in a nanotube. Because the liquid is heterogeneous, the static part of the
liquid stress tensor is not scalar and the equations of hydrodynamics are
not immediately valid. However, the results obtained for viscous flows can
be adapted at nanoscales \cite{Landau}. We denote the velocities by $%
\boldsymbol{v}_i=(0,0,w_{i})$ where $w_{i}$ is the velocity of component $i$
of the fluid mixture along the nanotube axis. When we neglect the external
forces, as gravity force, the two-liquid mixture motions of the components
verify Eqs. \eqref{motion}. Simple fluids slip on a solid wall only at a
molecular level and in classical conditions, the kinetic condition at solid
wall is the adherence condition $(z=R \Rightarrow   w_{i}=0)$ \cite{Churaev}%
. Recent papers in non-equilibrium molecular dynamic simulations of three
dimensional micro-Poiseuille flows in Knudsen regime reconsider
micro-channels: the influence of surface roughness, surface wetting and wall
density are investigated. The results point out that the no-slip condition
can be observed for Knudsen flow when the surface is rough; the surface
wetting also substantially influences the velocity profiles \cite{Tabeling}.
But for smooth surfaces, as carbon nanotube walls, a slip velocity is the
key for characterizing the flows. With water flowing through hydrophobic
thin capillaries, there are some qualitative evidences for slippage \cite%
{Blake}. De Gennes \cite%
{De Gennes} said: \emph{the results are unexpected and stimulating and led
us to think about unusual processes which could take place near a wall}.
Experimental results are connected with the thickness of the film when the
thickness is of an order of the mean free path \cite{secchi1}. When the
free mean path, denoted $\ell_{path}$,
 is smaller than diameter $2R$ of
carbon nanotubes, the Knudsen number, denoted \emph{Kn}, is smaller than 1.
That is the case for a liquid where the mean free path is of the same order
than molecular diameters. For example in the case of liquid-water, \emph{Kn}
ranges between $0.5$ and $10^{-2}$ while the nanotube radius ranges between $%
1$ nm and $50$ nm. Majunder  \emph{et al}   note that
\emph{slip lengths of micron order for their experiments with
nanometer size pores and the adherence boundary condition at a
surface, commonly employed with the Navier-Stokes equation, is
physically invalid}   \cite{Majumder}. A slip regime occurs and the boundary
condition must be changed to take account of the slippage at the
solid surfaces.  \newline We consider the simple case when
$\boldsymbol{a}_{1}=\boldsymbol{a}_{2}\equiv \boldsymbol{a}$ and
$\boldsymbol{v}_{1} =\boldsymbol{v}_{2} \equiv \boldsymbol{v} =
[0, 0, w]$, corresponding to the same velocity and acceleration of the two-fluid
mixture components. In fluid/wall slippage, the condition at the
solid wall is written as
\begin{equation}
w=L_{s}\frac{\partial w}{\partial r}\quad {\mathrm{at}}\quad r=R,
\label{slip}
\end{equation}%
where $L_{s}$ denotes the Navier-length \cite{Landau}. The Navier-length is
not independent of the thickness of the flow and may be as large
as a few microns for very thin films \cite{Tabeling,secchi2}. \newline
For carbon nanotubes, the dynamics of liquid flows is studied in the case of
non-rough nanotubes. Equations \eqref{motion} yield 
\begin{eqnarray}
&&\rho \,{\boldsymbol{a}}+\rho _{1}\,\nabla\, [\,\mu _{1}(\rho _{1})-\lambda
_{1}\,\Delta \rho _{1}\,-\lambda _{3}\,\Delta \rho _{2}\,]  \notag \\
&& +\ \rho _{2}\,\nabla\,[\,\mu _{2}(\rho _{2})-\lambda _{3}\,\Delta \rho
_{1}\,-\lambda _{2}\,\Delta \rho _{2}\,] = \nu \,\Delta {\boldsymbol{v}},
\label{motion3}
\end{eqnarray}%
where $\rho =\rho _{1}+\rho _{2}$ and we define the dynamic viscosity of the
fluid mixture $\nu$ as $\rho\,\nu =\rho _{1} \nu_{1}+\rho _{1}\nu_{2}$; $\nu$
is assumed constant. Consequently, we consider the case when:\newline
\emph{(i)} the boundary conditions take account of the slip condition %
\eqref{slip}, \newline
\emph{(ii)} the liquid nanoflow thickness is small with respect to
transverse dimensions of the wall and $2R\ll \ell_{path} $, \newline
\emph{(iii)} the flow is laminar corresponding to a velocity component along
the wall large with respect to the normal velocity component to the wall
which is negligible,\newline
\emph{(iv)} the permanent velocity vector $\boldsymbol{v}$ varies along the
direction orthogonal to the wall, $\nabla \rho $ is normal to $\boldsymbol{v}$ and the equation of continuity reads
\begin{equation*}
\rho \ \,\text{div}\, \boldsymbol{v} =0.
\end{equation*}%
Then, the density is constant along each stream line and the trajectories
are drawn on iso-density surfaces where $w=w(r)$.\newline
Due to the geometry, for permanent motions, the acceleration is null.
Equations of motion separate as follows \newline
$\bullet \ \ $ Along the $z$-coordinate, Eq. \eqref{motion3} yields
\begin{equation}
\begin{split}
\frac{\partial p}{\partial z}=\nu \,\Delta \,w \quad \mathrm{with}\quad
\Delta \,w=\frac{1}{r}\frac{d}{dr}\left( r\ \frac{dw}{dr}\right)\ \,
\mathrm{and}\ \, p=p_1+p_2,  \label{axequ}
\end{split}
\end{equation}
where $p_1$ and $p_2$ are the partial pressures of the two components.%
\newline

\indent $\bullet \ \ $ In the plane orthogonal to the tube axis, Eqs. %
\eqref{motion} yield
\begin{align}
\begin{split}
& \frac{\partial }{\partial r}\,(\mu _{1}(\rho_1)-\lambda_1 \,\Delta \rho_1
\,-\lambda_3 \,\Delta \rho_2)=0,\\
&  \frac{\partial }{\partial r}\,(\mu
_{2}(\rho_2)-\lambda_3 \,\Delta \rho_1 \,-\lambda_2 \,\Delta \rho_2)=0.
\label{tgpart}
\end{split}
\end{align}%
{It is fundamental to note that Eqs (\ref{tgpart}) yield  the same equations as at equilibrium. This result is the key of the distribution of densities and volume proportion between water and ethanol. Equation \eqref{axequ} is written as}
\begin{equation}
\frac{1}{r}\frac{d}{dr}\left( r\ \frac{dw}{dr}\right) =-\frac{\wp }{\nu}\,,
\label{axequ1}
\end{equation}%
where $\wp $ denotes the pressure gradient along the nanotube. The
cylindrical symmetry of the nanotube yields the solution of Eq. (\ref{axequ1}%
) in the form
\begin{equation*}
w=-\frac{\wp }{\nu }\,\frac{r^{2}}{4}+b,
\end{equation*}
where $b$ is constant. Condition (\ref{slip}) implies
\begin{equation*}
-\frac{\wp }{4\nu}R^{2}+b=L_{s}\frac{\wp }{2\nu }R
\end{equation*}%
and consequently,
\begin{equation*}
w=\frac{\wp }{4\nu }\left( -r^{2}+R(R+L_{s})\right).
\end{equation*}%
For liquids, the density in the nanotube is closely equal to $\rho _{\ell b}=
\rho_{1\ell b}+\rho_{2\ell b}$ and the volume flow through the nanotube is 
\begin{equation*}
Q=2\pi \displaystyle\int_{0}^{R}w\,r\,dr=\frac{\pi\,\wp }{8\,\nu }%
\,R^{3}(R+4L_{s}).
\end{equation*}
$Q_{o}$ denoting the Poiseuille flow corresponding
to a tube of the same radius $R$,  
\begin{equation}
Q=Q_{o}\left( 1+\frac{4L_{s}}{R}\right).  \label{flow}
\end{equation}%
In most cases, the Navier length is of the micron order ($L_{s}=1\,\mu
m=10^{3}\,$nm) \cite{Majumder}. If we consider a nanotube with $R=2\,$nm, we
obtain $Q=2\times 10^{3}\,Q_{o}$. For $R=50\,$nm, corresponding to the upper
limit nanotube-radius for nanofluidics with respect to microfluidics, we
obtain $Q=40\,Q_{o}$. Consequently, the flow of liquid in nanofluidics is
drastically more important than the Poiseuille flow in cylindrical tubes.
\newline
When the mixture mother-bulk is vapor, in the carbon nanotube the phase is
generally liquid and the volume flow through the nanotube is approximately
\begin{equation*}
Q=Q_{o}\frac{\rho _{\ell}}{\rho _{v}}\left( 1+\frac{4L_{s}}{R}\right).
\label{flow2}
\end{equation*}%
In the case of water and ethanol we have $\rho _{\ell}/ \rho _{v}\sim 10^2$ to $%
10^3$ and we get a volume flow at least $10^{2}$ time more important than the
volume flow obtained by Eq. (\ref{flow}). Then
\begin{equation*}
Q \sim 10^{5}\,Q_{o}\,.
\end{equation*}
The extremely large slip lengths measured in carbon nanotubes greatly reduce
the fluid resistance and nanoscale  structures could create extraordinarily
fast flows as it is in biological cellular channels \cite{Bonthuis}.

\section{Conclusions}

{In this paper, we propose a continuous model for the profile of densities between the two components of a fluid mixture. The model corresponding to the mean-field theory with hard-sphere molecules provides a free energy in the fluid mixture and an energy on the nanotube wall which yield a system  of second order differential equations   for equilibrium and for dissipative motions. Depending on  the wall's quality associated with the coefficients presented in Eq. \eqref{surfacenergy}, we obtain boundary conditions that permit to integrate the system. Thereby, carbon cylindrical  nanotubes with diameters ranging from 1 to 100 nm  can be studied. The carbon nanotubes are
filled up with liquid mixture of water-ethanol in a volume proportion imposed by the external mixture bulk.  For nanotubes of larger diameters essentially corresponding to microtubes, we obtain that the fluid phases can be liquid or vapor
according to the chemical properties of the tube walls and mother bulks with possible liquid-vapor interfaces. 
For tubes with small diameters, essentially corresponding to nanotubes, the fluid flows are liquid mixtures and can be significantly
greater than usual Poiseuille's flows, especially if the mixture
mother-bulks consist of vapor. \newline 
To obtain realistic densities of liquid ethanol and liquid water, we limited to three cases of concentration ($ c = 0.3, 0.5, 0.7$). The cases where $c$ is close to $0$ or $1$ correspond to vapor presence of one of the constituents; thus, the composition of the mixture must be quite far from these cases. It might be interesting to take up the problem with densities for which one of the components corresponds to a vapor density but in this case, Eqs. (16) will have to be modified.\newline For small nanotube
diameters, we notice that, in
the carbon nanotube, the ratio between water and ethanol  increases  
in favor of ethanol.  These results,
obtained by using a nonlinear model of continuum mechanics, are in
good agreement with the simulations of molecular dynamics
\cite{simulation}. Recent experiments confirm these results
\cite{secchi1,secchi2} and could be experimentally measured by indirect methods as Landau-squire plume measurements \cite{Landau,secchi3}.}

% \vspace{0.5cm}
\noindent
{\small \emph{\textbf{Acknowledgements :}}  {\small H.G. thanks  National Group of Mathematical Physics GNFM-INdAM for its support as visiting professor at  {\small{the Department of Mathematics, University of Bologna.\newline {\small This work was partially  supported  by National Group of Mathematical Physics GNFM-INdAM (A.M. and T.R.)}.}}
}}

\end{document}